\documentclass[12pt]{article}
\usepackage{titling}
\usepackage{mathpazo}
\usepackage[T1]{fontenc}
\usepackage[utf8]{inputenc}

\usepackage{dcolumn}
\usepackage{bm}
\usepackage{braket}
\usepackage{tabularx}
 \usepackage{booktabs}
 \usepackage{enumitem}

\usepackage{xcolor}
\usepackage{amsmath}
\usepackage{bm}
\usepackage{relsize}
\usepackage{comment}
\usepackage{graphicx}
\usepackage[UKenglish]{babel}
\usepackage{microtype}                      
\usepackage{xcolor} 
\usepackage{hyperref}                   	
\usepackage{lipsum}
\usepackage{fancyhdr}
\usepackage{picture} 
\hypersetup{								
	colorlinks=true,      		            
	linkcolor=blue,                        	
	filecolor=blue,                     	
	citecolor=blue,    	        			
	urlcolor=blue,                 	   		
	pdffitwindow=false,               	 	
	pdfstartview={FitH},               		
	pdfauthor={Huw Price},           		
}

\usepackage{braket}
\usepackage{physics}
\usepackage{todonotes}
\usepackage{enumitem}

\newcommand{\upV}{$\wedge$}

\usepackage{tikz}
\usetikzlibrary{shapes,decorations,arrows,calc,arrows.meta,fit,positioning}
\tikzset{
    -Latex,auto,node distance =1 cm and 1 cm,semithick,
    state/.style ={ellipse, draw, minimum width = 0.7 cm},
    point/.style = {circle, draw, inner sep=0.04cm,fill,node contents={}},
    bidirected/.style={Latex-Latex,dashed},
    el/.style = {inner sep=2pt, align=left, sloped}
}

\usepackage{rotating,stackengine,scalerel}
\newcommand\wye{\scalerel*{\stackengine{-1pt}{%
  \rotatebox[origin=c]{30}{\rule{10pt}{.9pt}}\kern-1pt%
  \rotatebox[origin=c]{-30}{\rule{10pt}{1.3pt}}}{%
  \rule{.9pt}{10pt}}{O}{c}{F}{F}{S}}{\Delta}}

\begin{document}

\hypertarget{taming-entanglement}{%
\title{Taming Entanglement}\label{bell}}
\author{Huw Price\thanks{Trinity College, Cambridge, UK; email \href{mailto:hp331@cam.ac.uk}{hp331@cam.ac.uk}.} {\ and} Ken Wharton\thanks{Department of Physics and Astronomy, San Jos\'{e} State University, San Jos\'{e}, CA 95192-0106, USA; email \href{mailto:kenneth.wharton@sjsu.edu}{kenneth.wharton@sjsu.edu}.}}
\date{\today}

\begin{titlingpage}
    \maketitle
   \begin{abstract}
\noindent  
In statistics and causal modeling it is common for a selection process to induce correlations in a subset of an uncorrelated ensemble.  We propose that EPR and Bell correlations are selection artefacts of this kind. The selection process is preparation of the initial state of the relevant experiments. Choice of initial state amounts to preselection of a subensemble of a larger, uncorrelated, virtual ensemble of possible histories.  Because it is preselection rather than postselection, the resulting correlations support the intuitive counterfactuals of the EPR argument and Bell nonlocality. In this respect, and in its temporal orientation, the case differs from familiar forms of selection bias. Given the ubiquity of quantum entanglement, the result may  thus be of independent interest to students of causal modeling. The paper concludes with a discussion of its novel implications in that field.\\

\end{abstract}
\end{titlingpage}


\section{The origins of entanglement}\label{sec:origins}

Quantum entanglement was first clearly identified, and named, by Erwin Schrödinger in 1935  \cite{Sch35a,Sch35b}. Schrödinger was responding to a now-famous paper by Einstein, Podolsky and Rosen (EPR) \cite{EPR}. He notes that in the two-part quantum systems that EPR discuss, two components that have just interacted cannot be described independently in the way that classical physics would have allowed. As he says:
\begin{quote}
When two separated bodies that each are maximally known come to interact, and then separate again, then such an \textit{entanglement} of knowledge often happens. \cite[§10]{Sch35a} 
\end{quote}
Schrödinger emphasises the centrality of this point to the new quantum theory.
\begin {quote}
I would not call that \textit{one} but rather \textit{the} characteristic trait of quantum  mechanics, the one that enforces its entire departure from classical lines of thought. \cite[555]{Sch35b} 
\end{quote}

More recently, entanglement has been described as
‘the essential fact of quantum mechanics’ \cite{Suss14}, and ‘perhaps its weirdest feature’ \cite{Weinberg13}.  Roger Penrose says that entanglement presents `two quite distinct mysteries’:
\begin{quote}
The first mystery is the phenomenon itself. How are we to come to terms with quantum entanglement and \emph{to make sense of it in terms of ideas that we can comprehend,} so that we can manage to accept it as something that forms an important part of the workings of our actual universe?  \cite[591, emphasis added]{Penrose04} 
\end{quote}
As for the second mystery, Penrose puts it like this: 
\begin{quote}
Since, according to quantum mechanics, entanglement is such a ubiquitous phenomenon
\ldots\ 
why is it something that we barely notice in our direct experience of the world? Why do these ubiquitous effects of entanglement not confront us at every turn? \cite[591]{Penrose04}
\end{quote}
Penrose says that this second issue has not `received nearly the attention it deserves, people’s puzzlement having been \ldots\ concentrated on the first.’ \cite[591]{Penrose04}

Penrose’s two mysteries provide a useful framing for the present argument. Our main proposal throws light on the first, as we interpret it. We explain how the remote correlations of entanglement may have a familiar origin. In the terminology of statistics and causal modeling (which we explain below), they may be a special sort of \textit{selection artefact.}
  For simplicity, we present the argument initially in terms of the kind of correlations discussed by EPR in 1935. It will then be easy to see that it also works for the  more subtle correlations discovered by John Bell in the 1960s \cite{Bell64}. 
 
 As for Penrose's second mystery, it highlights an  issue \textit{not} addressed by the present proposal. If entanglement correlations are selection artefacts, why are they confined to the quantum realm? At the end of the paper (\S\ref{sec:discussion}.4), we ask what this issue looks like, in the light of our proposal for the first mystery.

 The proposal has some novel implications for causal modeling and statistics. We close with a brief discussion of these consequences (\S\ref{sec:discussion}.5). The present approach may have been missed, in part, because it does require amendments to the familiar map in these ways.

\section{The EPR-Bohm experiment}\label{sec:epr-bohm}

We will use a version of the EPR experiment proposed by David Bohm \cite{Bohm51}, and later discussed by Bell. In this experiment, a source produces pairs of entangled spin-\textonehalf\ particles, jointly described by a certain quantum state. 
The particles go to measuring devices at remote locations A and B, which measure their spin either in direction X or direction Y. 

Adapted to this case, the crucial feature noted by EPR and Schrödinger is that the results at A and B are correlated: if the same measurement is made at both, they always yield the same result.\footnote{This is assuming a particular initial state corresponding to `parallel spins'.} 
There are two possible results, which for convenience we’ll denote below by 0 and 1. On each side, the string of 0s and 1s that we get from our measurements look entirely random. But they are perfectly correlated, always the same at A and B when the same measurement is made on both sides. 

Note that this correlation is \textit{counterfactually robust,} in the following sense. If the actual result was (say) 0 on both sides, then, holding fixed the measurement settings on both sides, we know that if the result at A \textit{had been} 1, then the result at B would also have been 1.\footnote{Some writers deny that such counterfactuals make sense in QM, often for reasons related to EPR and Bell (see \cite{Muynck94, Hess16}, for example). We set such views aside here,  taking for granted the intuitive counterfactuals of the EPR case, and proposing an account of their origins. More on the Bell case in \S\ref{sec:tobell} below.\label{fn_1}}

EPR and Schrödinger concluded that there must be some property of the two particles, fixed in advance, to produce this correlation: in modern terminology, the correlated measurement results must have a \textit{common cause.} Since such a common cause is not present in the standard quantum description, they concluded that that description is incomplete. They realised that a common cause would be unnecessary if the measurement at A could directly influence the result of that at B, or vice versa, but they thought that that was absurd. The systems could be arbitrarily far apart, and as Schrödinger put it, ‘[m]easurements on separated systems cannot affect one another directly, that would be magic’ \cite[§12]{Sch35a}.

In the 1960s, however, Bell showed that under plausible assumptions, the common cause proposal is incompatible with the predictions of quantum mechanics (QM) in more general versions of this experiment, allowing additional possible  measurement settings. The relevant quantum predictions have since been comprehensively confirmed, and Bell’s work is often interpreted as showing that QM does involve some sort of action at a distance, of the sort that EPR and Schrödinger took to be absurd. Bell himself took it this way, in his initial paper. Referring to EPR’s viewpoint, he summarises his own argument like this:
\begin{quote}
In this note that idea [of EPR’s] will be … shown to be incompatible with the statistical predictions of quantum mechanics. It is the requirement of locality, or more precisely \textit{that the result of a measurement on one system be unaffected by operations on a distant system with which it has interacted in the past,} that creates the essential difficulty. \cite[195, emphasis added]{Bell64}
\end{quote}
There are other views of the consequences of Bell’s work and the relevant experiments, and we will not attempt a survey here.\footnote{See \cite{Myrvold21} for discussion. Bell's own view evolved somewhat, in  later work, as these authors note.} The present proposal is novel, so far as we know. 

Although the central idea is unfamiliar in this context, it is standard fare in statistics and causal modeling. There, it is well-known that there is a {\textit{third option} for explaining correlations, in addition to common causes and direct causal influences. Some correlations are \textit{selection artefacts,} manifestations of so-called \textit{selection bias.}   One familiar example 
is \textit{survivorship bias}  \cite{Czeisler21}. 

The main novelty of the present proposal is to use this third option to account for entanglement correlations. It is easy to find a relevant selection mechanism. As we shall see, preparation of the initial state of the experiment does the trick.\footnote{We are indebted here to Gerard Milburn, who first proposed that we think of preparation as selection. For some readers ‘preparation’ may suggest common causes. We stress that that is not what we have in mind. We mean the simple operational notion common to everyone who discusses quantum experiments, whatever their take (or lack of one) on quantum ontology. See \S\ref{sec:toy}.3 for further discussion.} The third option then allows a simple explanation of EPR and Bell correlations: they are artefacts of these (pre)selections. By analogy with survivor bias, they may be regarded as `predecessor bias'. 

This proposal is certainly couched ‘in terms of ideas that we can comprehend’, as Penrose puts it. All the ingredients are familiar.\footnote{Or nearly so. There are some novelties by the lights of causal modeling. See \S\ref{sec:simpson}, \S\ref{sec:discussion}.5.} But the observation that the recipe applies in these cases seems to be new.\footnote{Three notes. First, it is simple to show \textit{that} the recipe applies in these cases, but harder to say \textit{why} it applies (and not, apparently, in classical cases). That's Penrose's second mystery; see \S\ref{sec:discussion}.4. Second, some readers may feel that the point is \textit{too} simple, once noticed, and doubt that it can tell us anything new about entanglement; we reply to this in  \S\ref{sec:discussion}.2. Third, the claim that Bell correlations are selection artefacts is not completely new. It has been has been made by several writers for a small class of cases involving delayed-choice entanglement-swapping; we discuss those cases and their relevance to the present proposal in \S\ref{sec:does}. It has also been proposed by \cite{Muller20}, in an interpretational framework Müller calls Algorithmic Idealism; thanks to Markus Müller here.\label{fn5}}

The next section outlines the basic proposal, for the EPR case. We begin with a toy model based on the EPR-Bohm experiment. It is an urn model, in the sense familiar in many fields \cite{John77}. As such, though we base its structure  on the actual quantum experiment, it is not itself a quantum system. However, it illustrates key features of the present proposal, in a way which readily maps over to the real quantum case.  Again, we focus for now on EPR correlations, though the extension to Bell correlations  will be straightforward (\S\ref{sec:tobell}).  

\section{Outline of the proposal}\label{sec:toy}
\subsection{An EPR-Bohm urn model}
The EPR-Bohm experiment may be conducted with `antiparallel' rather than `parallel' spins. It then produces \textit{anticorrelated} results, when the same measurement is made on both sides. The results at A and B are guaranteed to be \textit{different,} rather than guaranteed to be the \textit{same.}

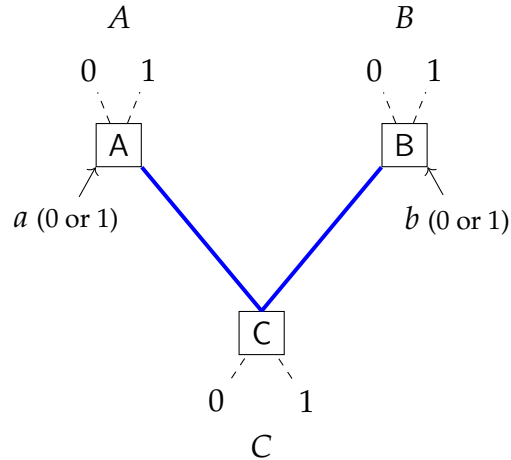
\begin{figure}[t]
\centering
\begin{tikzpicture}
    \node[draw,rectangle,minimum width=0.6cm,minimum height=0.5cm] (I) at (3.1,-2.5) {\textsf{C}};
\node[] (I0) at (2.5,-3.4) {0};
 \node[] (I3) at (3.7,-3.4) {1};
 \node[] (C) at (3.1,-4) {$C$};

  \path [dashed,-] (I0) edge (I);
  \path [dashed,-] (I3) edge (I);

    \node[draw,rectangle,minimum width=0.6cm,minimum height=0.5cm] (Abox) at (1.2,0) {\textsf{A}};
   \node[] (Alabel) at (1.2,1.7) {$A$};
    \node[] (A0) at (0.8,1) {0};
    \node[] (A1) at (1.6,1) {1};
    \node[] (a) at (0.5,-1) {$a$ \footnotesize{(0 or 1)}};
    \node[draw,rectangle,minimum width=0.6cm,minimum height=0.5cm] (Bbox) at (5,0) {\textsf{B}};
     \node[] (Blabel) at (5,1.7) {$B$};
       \node[] (B0) at (4.6,1) {0};
    \node[] (B1) at (5.4,1) {1};
    \node[] (b) at (5.7,-1) {$b$ \footnotesize{(0 or 1)}};

   \path [dashed,-] (Abox) edge (A0);
   \path [dashed,-] (Abox) edge (A1);
   \path [->] (a) edge (Abox.south west);
    \path [dashed,-] (Bbox) edge (B0);
   \path [dashed,-] (Bbox) edge (B1);
    \path [->] (b) edge (Bbox.south east);

   \draw [color=blue,line width=1.5pt,-] (I.north) -- (Abox.south east);
   \draw [color=blue,line width=1.5pt,-] (I.north) -- (Bbox.south west);

\end{tikzpicture}
\caption{EPR-Bohm experiment with random inputs at C} \label{fig:urn}
\end{figure}

Imagine a device combining both kinds of experiment. Pairs of entangled spin-\textonehalf\ particles are produced at random with parallel or antiparallel spins. The particles proceed to measuring devices at A and B, which measure their spin either in direction X ($0$)  or Y ($1$). This measurement setting is again chosen at random, and independently at A and B. Finally, an outcome is recorded at A and at B – this may be either 0 or 1. (See Figure~\ref{fig:urn}.)

Each run of this device can be described by five binary bits: one  for the  initial state ($C$), and two each  for the two settings ($a$,$b$) and two outcomes ($A$,$B$). Suppose that these bits are recorded at the corners of a pentagonal tile, as in Figure~\ref{fig:pent}. (The black dot orients the tile, by identifying the $C$ corner.) Let these five-bit tiles be collected in an urn, with no record of the order in which they arrive; and imagine that tiles are drawn from the urn at random.

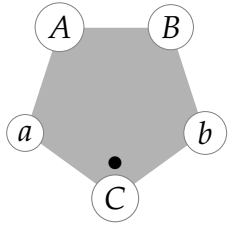
\begin{figure}[bt]
\centering
\begin{tikzpicture}

\node[regular polygon, fill=gray!60, regular polygon sides=5, minimum size=2.5cm, rotate=180] (a) {};
    \node[ inner sep=2pt, circle, draw=gray,  fill=white] at (a.corner 1) (c) {$C$};
     \node[inner sep=2pt, circle, draw=gray, fill=white, very thin] at (a.corner 2) {$b$};
     \node[inner sep=2pt, circle, draw=gray, fill=white, very thin] at (a.corner 3) {$B$};
      \node[inner sep=2pt, circle, draw=gray, fill=white, very thin] at (a.corner 4) {$A$};
      \node[inner sep=2pt, circle, draw=gray, fill=white, very thin] at (a.corner 5) {$a$};
     \node (f) at (0,-0.8) {$\bullet$};

\end{tikzpicture}
\caption{The pentagonal tile} \label{fig:pent}
\end{figure}

It is easy to confirm that there are no correlations between the A-side values and the B-side values shown on the tiles, as they emerge from the urn. The EPR correlations have been washed out by the random mix of correlated and anticorrelated cases. Figure~\ref{fig:urntab} shows the relative frequencies in the $C=0$ (correlated, left) and $C=1$ (anticorrelated, right) cases. Here ‘00’ means ‘Setting 0, outcome 0’, etc. The A-side possibilities are shown across the top, and the B-side possibilities down the side. It is easy to see that when we average each cell over the two tables, the resulting table shows no correlations at all.

\begin{figure}[hbt]
\centering

\includegraphics[width=10.5cm]{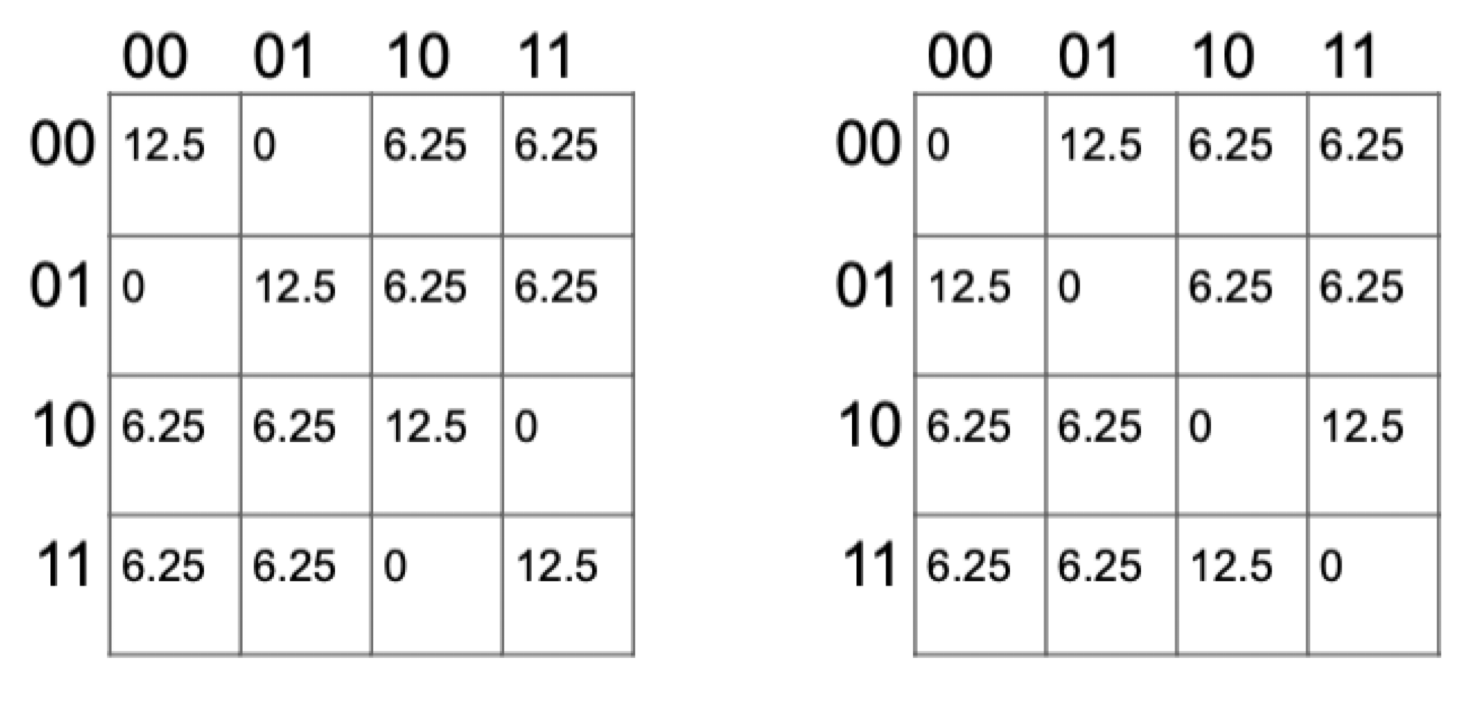}
 
\caption{Frequencies for the EPR-Bohm urn model} \label{fig:urntab}
\end{figure}

Suppose that a particular tile shows identical A and B settings and identical A and B outcomes. Let's ask a counterfactual question. If instead we'd drawn a tile with the same settings but a \textit{different} A outcome, would we also have found a different B outcome? No, because there is another possibility, equally likely. We could instead have found a tile with a different bit at C, in which case the A and B outcomes would also be different.

Still, we know that EPR correlations are in there somewhere. We can recover them by `postselection' – by discarding tiles showing (say) an anticorrelated initial state ($C=1$), retaining those showing a correlated initial state. In this postselected ensemble – the ‘retain’ pile – the tiles display familiar EPR correlations, as in the left hand table in Figure~\ref{fig:urntab}. If a tile shows the same A and B settings, it also shows the same A and B outcomes. 

Let’s ask a counterfactual question again. Choose a particular tile in the retain pile. Suppose for definiteness that it was the seventeenth tile drawn from the urn, and that it does display identical A and B settings, and hence identical A and B outcomes. 

Suppose that this tile had shown the same settings but a different A outcome. Would it also have shown a different B outcome? Not necessarily. As before, there is another possibility. The seventeenth tile might have shown $C=1$, in which case it would have been added to the discard pile, not the retain pile. So the retain pile displays EPR correlations, but these correlations don’t support the kind of counterfactuals we expect in EPR cases. The correlations are selection bias, as people say. (Again, we explain this terminology below.)

Sharp-eyed readers may have noted that the case does support a different counterfactual. The seventeenth tile drawn from the urn is a record of a run of the experiment in Figure~\ref{fig:urn}. Referring to \textit{that particular run,} it does seem true to say that if the settings had been the same but the A outcome different, the B outcome would also have been different. In assessing that counterfactual, it seems natural to hold fixed the actual initial state (which, by assumption, was $C=0$). We’ll come back to this point below (\S\ref{sec:time-asymm}).

Returning to our toy model, we observed that postselecting the $C=0$ cases gave us EPR correlations, but not the counterfactuals typical of actual EPR experiments.  
But now let’s try a different way of getting rid of the $C=1$ tiles. Instead of \textit{postselecting}, let’s \textit{preselect.} Imagine we could restrict the  contents of the urn in advance, so that there were no $C=1$ tiles.  Every tile drawn from the urn would then be $C=0$, with no postselection required.  
So if there is some way to impose this preconstraint, the counterfactuals will work the way we want: the seventeenth tile couldn’t have been  $C=1$, because there were no such tiles in the urn. 

As we have described the model, however, this preconstraint is easy. Tiles are generated by the runs of the experiment in Figure~\ref{fig:urn}, in which C is a preparation variable, easily put under experimental control. With $C=0$ imposed by an experimenter there, every tile in the urn will also show $C=0$.

So in this `preconstrained' version of the urn model, we get regular, counterfactual-supporting EPR correlations. What’s interesting is the way we got there. We started with an ensemble with no such correlations, and produced them by throwing some of the ensemble away. For counterfactual robustness it was important that we do this proactively rather than retrospectively, but preselection didn’t require any extraordinary trick. As the model was described, it just required an ordinary, unremarkable ability to set the initial conditions of the urn-generating experiment.

For clarity, we stress again that the urn model is not a quantum system. We  imagined using a real EPR experiment to populate it with the frequencies we wanted, but that step, though helpful below, is inessential. We could have used other stochastic devices (e.g., tossed coins) as a source of randomness. The rules we need to impose are  simple. Tiles in the urn must satisfy $A+B+C=0  \pmod{2}$ when $a=b$ -- this gives us correlation or anticorrelation between $A$ and $B$, depending on the value of $C$ -- and the distribution should otherwise be random  on all variables. 

Despite being simple and classical, the urn model illustrates two important points:\begin{enumerate}
    \item It is easy to embed an EPR-correlated ensemble in a larger, uncorrelated ensemble, so that the correlated subensemble emerges by selection. \item If we do this by preselection not postselection, we get the same intuitive counterfactuals as in real EPR experiments.\end{enumerate}
 The next step is to transfer these lessons to the real (quantum) world.

\subsection{To the real world}

With modest idealisation, any run of a real EPR experiment of the kind shown in Figure~\ref{fig:urn} can be regarded as defining a five-cornered region in spacetime, with the physical operations corresponding to the variables $C,a,A,b,B$ at the five vertices. Each operation has a binary value, and the world itself may thus be regarded as a single huge urn, containing many five-bit spacetime tiles of this kind.\footnote{What counts as the \textit{world,} for these purposes? That's a good question, to which we'll return in \S\ref{sec:ur}, when we have a suitable answer on the table.}

The lessons of the toy model now apply. There are no EPR correlations in the urn as a whole.\footnote{Again, what `as a whole' means will be clarified in \S\ref{sec:ur}.} But such correlations emerge in subensembles, when we select for a particular value of $C$. If we make this \textit{preselection,} as we do when we set the initial conditions of the experiments concerned, 
 then the resulting correlations are counterfactually robust, as in the toy model. As we'll see below (\S\ref{sec:tobell}), all of this carries over the Bell correlations, which can be explained in the same way, via a `universal' urn model of this kind.

 In the case of EPR correlations, however, this whole proposal may seem redundant. As we noted, it is easy to build a classical device to generate the correlations on which the toy model depends. We simply need coin tosses to supply randomness, and a rule for sending the results to A and B, as outcomes for their two possible measurements 
 (building in correlation or anticorrelation, as required). Once we understand the mechanism in such a gadget, we understand where the correlations are coming from. What role could there be for a proposal about selection artefacts? 

 For real EPR correlations we don't have such a mechanism, and Bell's work makes the search for one unpromising (at best). But doesn't the analogy with this classical EPR gadget suggest that our proposal is misguided, in not even seeking such a mechanism? 

\subsection{Two explanatory stances}

 This is an important challenge. To meet it, we need a distinction between two kinds of explanatory questions, or stances. The first  is broad and top-down. It treats the existence of correlations and (apparent) dependencies as anomalies, somewhat like symmetry-breaking. How can there be such structures in the world? The urn model illustrates how preselection of initial states provides an answer, in some cases. It generates counterfactually-robust correlations by `restriction' from a larger uncorrelated ensemble. Later, we'll see that this kind of answer has precedents elsewhere in physics. 

The second stance is narrower. It takes the initial state for granted, in effect, and asks how it is `connected' to the observed outcomes. It asks for a \textit{mechanism} for the connection, in a familiar use of that term; we can call it the \textit{Mechanistic Stance} (MS). A good name for the first approach is harder to find. We'll call it the \textit{Structural-Operational Stance} (SOS).\footnote{We will explain this label further in \S\ref{sec:discussion}.3, linking `structural'  to earlier uses of the term in quantum foundations.} 

The two stances ask different questions, and hence are not rivals, at least not directly. Both can give informative answers about the same case. This is true of the EPR gadget just described, though MS is simple here, while it would take some work to motivate an interest in SOS.\footnote{Here's a story to do the job. Imagine a factory producing our gadgets, in equal numbers of correlating and anticorrelating versions. Suppose this flat distribution is needed to prevent signalling, in some use of  the devices. 
Our codes are cracked, and we trace the problem to a bias in the factory. 
It has been hacked by our enemies, and is producing more correlating than anticorrelating gadgets. Here, we explain a (regrettable) lack of symmetry in the world, in terms largely insensitive to the nature of the mechanisms.} But in the real EPR case, as  Bell showed, MS is far from straightforward. 
SOS, on the other hand, turns out to be informative. It throws light on Penrose's first mystery, showing how we can make sense of EPR and Bell correlations `in terms of ideas that we can comprehend', as Penrose puts it. Do we still need MS? That's a good question, to which we'll return at the end (\S\ref{sec:discussion}.4); we may well need MS for Penrose's second mystery (\textit{inter alia}).

As we have said, the notion of a selection artefact is familiar in statistics and causal modeling. The next section introduces some terminology from those fields that will  help in explaining the present proposal further. We will also introduce some new terminology, needed where the proposal diverges from orthodox causal modeling. 

\section{Simpson, Berkson, and two kinds of fork}\label{sec:simpson}

Let’s begin with Simpson’s Paradox. The \textit{Stanford Encyclopedia of Philosophy} describes it like this.
\begin{quote}    
Simpson’s Paradox is a statistical phenomenon where an association between two variables in a population emerges, disappears or reverses when the population is divided into subpopulations. \cite{Sprenger21}
\end{quote}
It is easy to see that we have an example in Figure~\ref{fig:urntab}. There are no correlations when the two cases are combined, but EPR correlations emerge in the two subpopulations. Put like this, the idea is unremarkable. We could imagine much simpler cases. 

The interesting aspects of such cases often emerge from  \textit{selection effects.} These arise when the method by which a subpopulation is selected intersects with the kind of stratification into subpopulations that the general phenomenon describes. Confusion can arise if we don’t pay sufficient attention to the means of selection, and to the fact that correlations in the selected subpopulation may not mirror those in the population as a whole.

It is often helpful to describe such cases in the terminology of causal modeling. In this framework, a \textit{common cause} is an event, or variable, with two (or more) distinct effects, neither of which is a cause of the other. In the graphical format of Directed Acyclic Graphs (DAGs), Figure~\ref{fig:cc} shows a case in which F is a common cause of G and H. In a simple case of this kind G and H will be correlated because they have a common cause, but this correlation disappears if we ‘condition on’, or restrict our attention to, the cases in which the common cause variable takes a particular value. Conditioning on a common cause \textit{screens off} the correlation between its joint effects.

\begin{figure}
\centering
\begin{tikzpicture}
    \node (a) at (0,0) {G};
    \node (F) at (3,-3.3) {F};
    \node (b) at (6,0) {H};
\coordinate (coll) at (3,-3); 
  \path[color=gray,rounded corners,line width=2pt] (coll) edge (a);
   \path[color=gray,rounded corners,line width=2pt] (coll) edge (b);

\end{tikzpicture}
\caption{A common cause} \label{fig:cc}
\end{figure}
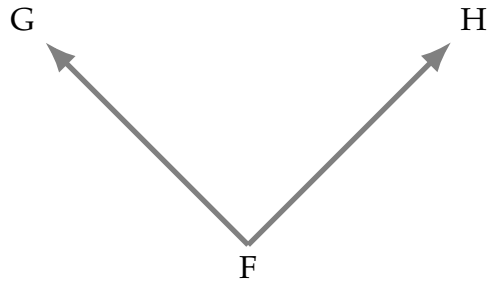

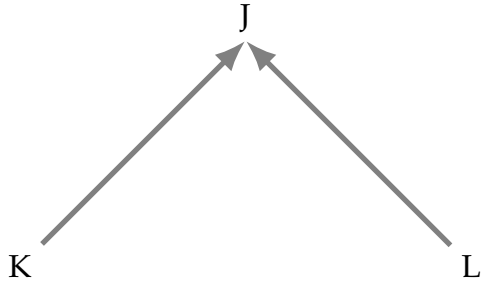
\begin{figure}
\centering
\begin{tikzpicture}
    \node (a) at (0,0) {K};
    \node (F) at (3,3.3) {J};
    \node (b) at (6,0) {L};
\coordinate (coll) at (3,3); 
  \path[color=gray,rounded corners,line width=2pt] (a) edge (coll);
   \path[color=gray,rounded corners,line width=2pt] (b) edge (coll);

\end{tikzpicture}
\caption{A common effect, or collider} \label{fig:ce}
\end{figure}

In the same framework, a \textit{common effect} is an event, or variable, with two (or more) distinct direct causes. Figure~\ref{fig:ce} shows a case in which J is a common effect of K and L. In these cases, conditioning works the other way. In the simple case shown, K and L are independent, or uncorrelated; but they may become correlated if we condition on the cases in which the common effect variable J takes a particular value. This is Simpson’s Paradox at work: K and L are uncorrelated in the general population, but correlated in subpopulations defined by the value of J.

Common effects are also called \textit{colliders,} for the obvious reason: they are variables in a DAG where two causal arrows collide. The correlation induced by conditioning on a collider is called \textit{collider bias.} The same phenomenon is also called \textit{Berkson’s bias,} or Berkson’s paradox. Berkson was a Mayo Clinic statistician and physician who identified the effect in the 1940s \cite{Berkson46}.\footnote{The point dates back at least to the Cambridge economist A C Pigou \cite{Pigou11}; thanks to Jason Grossman and George Davey Smith here. Another early source -- decades before Berkson, and in the same country -- is the influential geneticist Sewall Wright \cite{Wright21}; here we are grateful to Clark Glymour.} He described a case in which patients were hospitalised with a symptom with two possible causes. These causes might be independent in the general population, but among the hospitalised patients with the symptom in question, they were strongly correlated: if they didn’t have one disease, they must have the other. 

Unlike  `collider bias', the term `Berkson’s bias' doesn’t build in the terminology of causal models. This will be an advantage here. Causation is a subtle matter in the cases we want to discuss, and it will be helpful not to be forced to refer to it, to describe the statistical relationships we need.

The Wikipedia entry on Berkson’s paradox \cite{Wiki25} gives a simple non-causal example, which we adapt here. Suppose Alice and Bob each toss a fair coin, the results being tallied in a $2\times2$ table (Figure~\ref{fig:bb}). There are four possible outcomes on each trial, equally likely, and the A result and B result are independent. For example, the probability that Bob tosses a head, given that Alice tosses a head, is \textonehalf.

\begin{figure}[t]
\centering
\includegraphics[width=5cm]{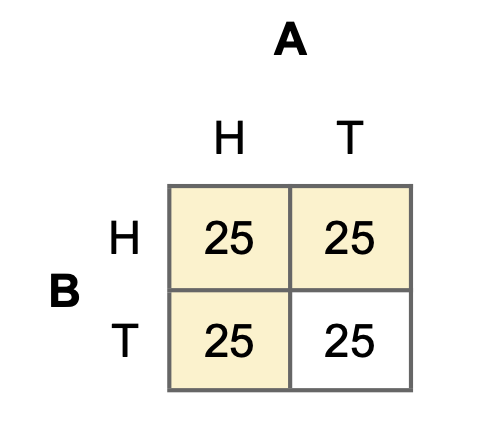}
\caption{Berkson's bias} \label{fig:bb}
\end{figure}

Now we restrict our attention to the cases in which there is at least one head. We thus ignore the lower right cell, where both results are tails. In the remaining (shaded) cells, P(A=H) = P(B=H) = 2/3, but P(A=H\&B=H) = 1/3. So, within this selected class, P(A=H\&B=H) < P(A=H).P(B=H). In other words, A=H and B=H are now negatively correlated. This correlation is a selection artefact: it appears because we have selected a particular subset of the class of all possible outcomes. 

We could represent this case as conditioning on a common effect, by treating ‘At least one head’ as a variable in a DAG, a common effect of A=H and B=H. But this isn’t necessary. The phenomenon we are interested in is present in the raw statistics, whether or not we give them a causal interpretation.\footnote{We could take the case even further from the causal realm by using a purely mathematical example, based, say, on frequencies of digits in the decimal expansion of $\pi$.} 

We noted that common causes and common effects are distinguished by how they behave under conditionalisation. Conditionalising on a common cause \textit{removes} correlations between the joint effects; conditionalising on a common effect (often) \textit{induces} correlations between its joint causes. This distinction will be crucial below, but the terminology is in some ways unhelpful. In particular, we want the freedom to discuss non-causal cases, such as the  example of Berkson's bias just given. 
Accordingly, we introduce the following terminology. We will call Figure~\ref{fig:ce}  a \textit{Correlating Fork} (CorrF) and Figure~\ref{fig:cc}  a \textit{DeCorrelating Fork} (DeCorrF). We stress that the labels refer to what happens under conditionalisation -- we have spared the reader that complexity in our choice of notation.\footnote{The use of the term `fork' for these structures derives from \cite{Reich56}. Our DeCorrF is what Reichenbach calls a \textit{conjunctive fork.} He wrongly claims that most common effects (Figure~\ref{fig:ce}) are conjunctive forks; see \cite{Sober92}.} 

As depicted in Figures~\ref{fig:cc} and \ref{fig:ce}, these forks have opposite temporal orientations. That won’t always be the case. The terminology permits cases that are not temporal at all, being defined on statistical ensembles of something other than spatiotemporal events.\footnote{For example, again, frequencies of digits in the expansion of $\pi$.} 
In the present context, our most interesting examples of CorrFs will be spatiotemporal, but they are inverted compared to the familiar common effects of Figure~\ref{fig:ce}. They have their vertex in the past, relative to the other two nodes, in other words. This is one of the two novel implications for causal modeling of the current discussion. Thanks to QM, CorrFs with this reverse temporal orientation are actually exceedingly common. (We discuss this further in \S\ref{sec:discussion}.5.)

Why are the correlations produced by conditioning on a CorrF called selection \textit{artefacts?} The terminology seems to capture two things. First, they are \textit{produced by} the selection, in the obvious way. They are correlations that appear when we restrict our attention to a particular subclass of the data. Second, within the causal modeling framework, they are not the real deal -- not genuine \textit{causal} correlations. We have distanced ourselves a little from the causal modeling framework, but we have invoked a similar distinction. For example,  we observed that in our toy model, the postselected EPR correlations do not support the counterfactuals associated with ordinary EPR experiments.

Our proposal introduces a new complexity, however. We have said that \textit{some} CorrFs do support counterfactuals. We get it from preselection, as opposed to postselection, in our toy model. So we admit a class of correlations that are selection artefacts in the first sense -- they are produced by a selection procedure -- but not, or not straightforwardly, in the second sense. We need not take a stand on whether correlations produced by such a CorrF deserve to be called \textit{causation,} but they are not \textit{mere} correlations, in one of the senses often contrasted with causation. This is the second novel consequence for causal modeling; again, we discuss this further in \S\ref{sec:discussion}.5.

We need some terminology to distinguish between these two kinds of CorrFs. Let's use a simple non-quantum example. Suppose that Ivy College selects students for intellectual ability or athletic ability, not requiring both, and that these traits are independent in the general population. Admission to Ivy is thus a common effect of two independent causes. If we restrict our attention to students who are admitted, the two traits are anticorrelated: if a student is not athletic they must be clever, and vice versa. Imagine that a particular student, Holly, gets straight As but can’t catch a ball. Would she have been more athletic if she had not been so clever? No. She simply wouldn’t have been admitted to Ivy in the first place.\footnote{We are grateful to George Davey Smith for this example.} 

Stretching our imaginations a little, we can adapt this case to illustrate the difference between CorrF-induced selection artefacts that do or do not support counterfactuals. Suppose that Holly has a fairy godmother (FG), who tells her as a child that she is `fated' to be admitted to Ivy College, one way or other. Assuming that Holly trusts FG, it will seem from her point of view that the CorrF does support counterfactuals. Whichever ability will actually get her into Ivy, the other one \textit{would have done so,} if the first had not been sufficient. Being a clever child, Holly might thus plan to guarantee her academic results by neglecting her track practice. 

The case is not realistic, of course, but it is easy enough to imagine.\footnote{Easy enough to be the main plot device in children's fiction \cite{Vernon15}. We are grateful to Holly Andersen here.} We'll call a CorrF of this kind -- one that does support counterfactuals --  a \textit{Constrained} Correlating Fork (ConCorrF). ConCorrFs will play a big role below, where they won't depend on magic. Ordinary preselection will do the job, as in the EPR-Bohm urn model.

\section{Application to EPR correlations}\label{sec:application}

Let's now apply this terminology to the EPR case. As we noted in \S\ref{sec:epr-bohm}, EPR and Schrödinger argued that EPR correlations require what would now be called common causes, missing in the standard quantum formalism. Bell agreed that the argument was a reasonable one, but showed, as he put it, that the `reasonable thing just doesn't work.'\footnote{John Bell, quoted in \cite[84]{Bernstein91}.} In slightly more complicated versions of these experiments, the common cause hypothesis implies a mathematical relation now known as Bell’s Inequality. And QM predicts that actual experiments will violate Bell’s Inequality, in some cases.

Why did EPR think that the QM correlations required a common cause? In effect, they were invoking what later came to be called the Principle of Common Cause (PCC) \cite{Cricky21}. The following informal version of PCC will do for our purposes \cite{Hofer2013}: 
\begin{quote}
The Common Cause Principle says that every correlation is either due to a direct causal effect linking the correlated entities, or is brought about by a third factor, a so-called common cause.    
\end{quote}
EPR and Schrödinger took it to be obvious that there could be no direct causation, in the cases in question, because the systems could be arbitrarily far apart. As we said, Bell’s work is often interpreted as showing that QM does involve some sort of action-at-a-distance of this kind.

As we noted, however, {there's a third option.} It is well known that if there is a process of selection in play, correlations may also arise as selection artefacts. In the present case, our discussion in \S\ref{sec:toy} shows what the selection might be. In setting up an EPR-Bohm experiment to produce parallel spins, rather than a random mix of the parallel and antiparallel cases, we make the selection needed to reveal EPR correlations in a subclass of a larger, uncorrelated ensemble. 
Because it is preselection rather than postselection, the resulting correlations support the counterfactuals characteristic of EPR cases. We thus have a ConCorrF in the picture, in other words (with no magic needed). 

As foreshadowed in \S\ref{sec:simpson}, the ConCorrF in question opens to the future, not the past. It has the opposite temporal orientation to  familiar CorrFs like Figure~\ref{fig:ce}. This helps to explain why the option has been missed, presumably. It is a familiar beast in a very unfamiliar location.\footnote{A familiar beast in being simply a CorrF, at any rate. The notion of a \textit{Constrained} CorrF is also unfamiliar, but it comes with the location in a natural way, as the toy model shows.} 
At least as developed here, it also involves an unfamiliar shift in explanatory stance, from Mechanistic (MS) to Structural-Operational (SOS) in the sense of \S\ref{sec:toy}.3. So it is not a direct competitor to the two options allowed by PCC. Nevertheless, as we'll see, it provides illumination where they seem to fail.

Changing the stance in this way may seem like a trick. A reader may feel we're not answering the question she was asking. That's true, in a sense, but we hope to persuade such readers that our understanding of entanglement benefits from the change.\footnote{In physics as in philosophy, it is hardly news that a change of  question may be just what the doctor ordered.} 

The proposal may also seem a sleight of hand in another way.  Our urn model was an artificial construction, deliberately chosen to wash out the EPR correlations in the larger ensemble. This involved both a stipulation of the additional part of that ensemble, and a choice of a measure to guarantee ‘washing out’. Does the argument work if the two possible initial states don't have equal probability? If not, where are we to find such convenient ensembles and measures in the real world? 

We want to approach these questions indirectly, describing another case in which failure to recognise the role played by control of initial conditions leads to confusion in QM. In this well-known case the confusion, surprisingly, is on the part of Roger Penrose. One of the advantages of the case for us is that many people have identified Penrose's mistake.

Another advantage is that the key to Penrose's mistake is a familiar point about the origins of time-asymmetry in our universe, one that Penrose himself has been influential in describing. Having this point in view will enable us to explain how the present proposal is a special case of something much more familiar.

\section{Penrose on quantum measurement}\label{sec:penrose}

Penrose argues that there is a fundamental time-asymmetry in QM, in the component of the theory relevant to measurement. In  \cite{Penrose89, Penrose04}  he discusses the case in Figure~\ref{fig:pen}, describing it as follows.

\begin{figure}[t]
\centering
\includegraphics[width=11cm]{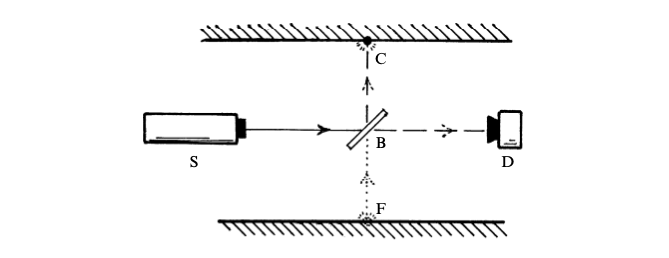}
\caption{Penrose on quantum measurement 
} \label{fig:pen}
\end{figure}

\begin{quote}
    A source S randomly emits single high-energy photons \ldots\ aimed at a beam-splitter B, tilted at 45$^\circ$ to the beam. If transmitted through B, the photon activates a detector D (route SBD); if reflected, it is absorbed at the ceiling C (route SBC). The quantum squared-modulus rule correctly predicts probabilities \textonehalf, \textonehalf. On the other hand, given that D registers, the photon could have come from S (route SBD) or from the floor F (route FBD). Used in the reversed-time direction, the squared-modulus rule incorrectly retrodicts probabilities \textonehalf, \textonehalf, which should be 1, 0. \cite[820]{Penrose04}
\end{quote}

A common response to Penrose's argument is described by John Baez \cite{Baez96}, who recounts presenting it to Penrose in person. 
\begin{quotation}
\noindent I told him I'd seen this argument and found it very annoying. He said that everyone said so, but nobody had refuted it. \ldots\ I said that \ldots\ if the whole system were in a box and had reached thermodynamic equilibrium, so the walls were just as hot as the lightbulb, we would be justified in concluding, upon seeing a photon at [D], that there was a 50-50 chance of it originating from [S] or [F]. It is only that our world is in a condition of generally increasing entropy that allows for the setup with a hot lightbulb and cool walls to occur, and we can't blame quantum mechanics for that time asymmetry.

[Penrose] thought a while and said that well, a condition of thermal disequilibrium was necessary for a measurement to be made at all, but the real mystery was why we feel confident in using quantum mechanics to predict and not retrodict. This mystery can be traced back to gravity, in that gravity is the root of the arrow of time. \ldots\ I was a bit disappointed that he didn't think my remarks dealt a crushing blow to this thought experiment, but I had to agree that if this was all he thought the moral of the experiment was, he was right.
\end{quotation}

Baez was right to be disappointed, in our view. The need to take into account the low entropy past in retrodiction is not some peculiar feature of QM. On the contrary, it is universal. Sean Carroll puts the point like this, for example:   
\begin{quote}
    Ordinarily, when we have some statistical system, we know some macroscopic facts about it but only have a probability distribution over the microscopic details. If our goal is to predict the future, it suffices to choose a distribution that is uniform in the Liouville measure given to us by classical mechanics (or its quantum analogue). If we want to reconstruct the past, in contrast, we need to conditionalize over trajectories that also started in a low-entropy past state. \cite{Carroll13}
\end{quote}

Interpreted as Penrose describes it to Baez, then, the point has nothing specifically to do with QM. Yet some years after that conversation, in \cite{Penrose04}, Penrose continues to present the point as a distinctive observation about QM. Referring to objections such as Baez's, he says:
\begin{quote}
Sometimes people have objected to this deduction, pointing out that I have failed to take into account all sorts of particular circumstances that pertain to my time-reversed description, such as the fact that Second Law of thermodynamics only works one way in time, or the fact that the temperature of the floor is much lower than that of the source, etc. But the wonderful feature of the quantum-mechanical squared-modulus law is that we never have to worry about what the particular circumstances might be! The miracle is that the quantum probabilities for future predictions arising in the measurement process do not seem to depend at all on considerations of particular temperatures or geometries or anything. If we know the amplitudes, then we can work out the future probabilities. All we need to know are the amplitudes. The situation is completely different for the probabilities for retrodiction. Then we do need to know all sorts of detailed things about the circumstances. The amplitudes alone are quite insufficient for computing past probabilities. \cite[821]{Penrose04}
\end{quote}
Again, however, the same is true of classical statistical probabilities, as Carroll's remark above makes clear. 

Penrose adds an endnote to this discussion, which takes us in the direction of our present concerns. 
\begin{quote}
I find it remarkable how much difficulty people often have with this argument. The matter is perhaps clarified if we contemplate numerous occurrences of \textit{this experiment,} taking place at various locations throughout spacetime. There are four alternative photon routes to be considered, SBD, SBC, FBD, and FBC. To see what the various probabilities are, we ask for the proportion of SBD, given S (forward-time situation), or for the proportion of SBD, given D (backward-time situation). The squared-modulus rule correctly gives the actual answer (50\%) in the first case, but it does not give the actual answer (nearly 100\%) in the second case. \cite[866, emphasis added]{Penrose04} 
\end{quote}
However, everything hangs on what we mean by an occurrence of \textit{this experiment.} If we stipulate that it means that the photon is at S at the beginning of the experiment, then we have put the asymmetry that Penrose wants in by hand, and it no surprise that we can extract it again later. But if we remove this stipulation for the reverse cases, stipulating only that the photon is at D at the end of the experiment, we get Baez's conclusion. So long as we also equalise for the thermodynamic factors -- which we might do, in principle, by imagining that the experiments in question occur as fluctuations in a universe in thermodynamic equilibrium overall -- then there is no asymmetry in the frequencies of the kind Penrose requires.

This example has several useful lessons for our discussion. For one thing, it illustrates how we have trouble taking account of the significance of ordinary control of initial conditions of experiments -- let's call this \textit{Initial Control} (IC), for short. Penrose's mistake -- even in his presentation in the last passage quoted above, supposedly generalised to `various locations throughout spacetime' -- is to ignore the role that IC is playing. 

An apt label for what Penrose is missing is \emph{predecessor bias}. The apparent time-asymmetry results from a restriction, imposed via IC,  on the \emph{history} of the systems in question.   The case shows that we can exhibit the role of predecessor bias by considering an imaginary regime in which IC is absent -- but that we need to do it properly, and not simply imagine a lot of cases in which it is present!

    The example also illustrates the explanatory benefits of considering this imaginary regime.  In Penrose's case, it is the key to avoiding a fallacious argument about a fundamental time-asymmetry in quantum measurement. In our case, it reveals how entanglement may be a product of predecessor bias, in the form of a ConCorrF. Ordinary IC does the constraining, but in order to see the role it is playing,  we need to imagine it absent.

The precise imaginative steps are somewhat different in the two cases, of course.  The key to avoiding Penrose's mistake is to see that from the perspective of the universe, as it were, there is equally likely to be a source at F on the floor of the laboratory. In effect, we need to hold fixed the quantum state responsible for photon emission, but to imagine it less tightly constrained in space.\footnote{It is still very tightly constrained, by the universe's standards, if the only two possible locations are S and F, but we don't need all the other options to make our point.}

Our proposal needs a different kind of generalisation. We need to imagine the \textit{initial state itself} less tightly constrained.\footnote{Or,  better -- to be faithful to our operational stance -- to imagine the initial \textit{preparation} less tightly constrained.} If we begin with the version of the EPR-Bohm experiment with parallel spins, then we need to add an equal mix of antiparallel spins to produce our larger, uncorrelated ensemble of results. 

We haven't yet offered a justification for thinking of this as an equally likely option, by the universe's lights, let alone described how it  generalises to other kinds of experiment. We'll come to those points below (\S\ref{sec:does}, \S\ref{sec:general}). But the last observation we can extract from the analogy with Penrose's argument is that equiprobability may be much more than we need. 

To see this, imagine that \textit{the universe itself} had a preference for producing photons at the wall rather than the floor of the kind of laboratory shown in Figure~\ref{fig:pen}. That would indeed give us some kind of time-asymmetry in the QM probabilities, its degree depending on the strength of the preference concerned. We don't need to rebut that rather bizarre possibility, to identify Penrose's mistake. Penrose isn't proposing that the universe has any such preference. He gets the asymmetry he claims to find by putting in by hand the fixed location of the source.

Similarly in our case. Even if it turned out that the universe itself had a preference for producing parallel over antiparallel spins, in cases of this kind, we would still have an SOS explanation for the resulting EPR correlations. They would arise because the universe itself would be doing the preselection, in favouring one class over the other.

This observation takes us in the direction of our next topic. It concerns a real and enormously significant case in which the universe does do something like this, favouring a tiny subset of a larger space of possibilities. We approach that topic if we ask where Initial Control comes from. We don't control the final conditions of experiments in the same direct way. Why does our universe have this striking time-asymmetry? The answer seems to lie in a more general story about the origins of most, perhaps all, of the time-asymmetric features of our physical universe. It will be useful in the remainder of the paper to have the key features of this account in view -- and here Penrose himself should be one of our guides.

\section{The origins of time-asymmetry}\label{sec:time-asymm}

You are walking across campus one windy day, and notice some scraps of paper, falling from a window. Picking up one scrap, you find some familiar words: ``There are more things in heaven and earth, Hora \ldots'' -- the printed words tail off at the torn edge of the paper. 

What should you expect if you pick up a nearby scrap, with a matching torn edge? It depends which building it is. If it is the primate lab, and you can hear the sound of typewriters, you might well expect gibberish. If it is the library, then more \textit{Hamlet,} or at any rate not gibberish.

The case illustrates one of the most important lessons of the physics of time-asymmetry, over the past 150 years. Josef Loschmidt was a  teacher and  colleague of Ludwig Boltzmann in nineteenth-century Vienna. In the 1870s he raised an objection to Boltzmann's attempt to explain in statistical terms why entropy always increases. Loschmidt pointed out that for every entropy-increasing possibility, there was a matching entropy-decreasing case, described by simply reversing the motions.\footnote{Loschmidt puts the crucial point like this: `Obviously, in every arbitrary system the course of events must be become retrograde when the velocities of all its elements are reversed' (\cite[139]{Lo76}, quoted in \cite{Uff24}).} 

So if statistics imply that entropy increases towards the future, shouldn't they also imply that it increases towards the past? This would mean that the apparent order around us, like the monkey's snippet of \textit{Hamlet,} is a random piece of organisation in a stream of gibberish. These days, the point is often made by imagining so-called Boltzmann Brains -- random, self-aware thinkers, coalescing by chance from the background chaos. 

The challenge to Boltzmann inspired by Loschmidt is known as the \textit{reversibility objection.} The objection itself, and the most influential present-day response to it, are here described by Mathias Frisch. 

\begin{quote}
If we assume an equiprobability distribution of micro-states compatible with a given
macro state of non-maximal entropy, then it can be made plausible that, intuitively, ‘most’ micro
states will evolve into states corresponding to macro states of higher entropy. However, if the
micro-dynamics governing the system is time-symmetric, then the same kind of considerations
also appear to show that, with overwhelming probability, the system evolved \textit{from} a state of
higher entropy. This undesirable retrodiction, which is at the core of the \textit{reversibility objection,}
can be blocked, if we {conditionalize} the distribution of micro-states not on the present macro-state but on a low-entropy initial state of the system. \ldots\ Albert and others argue that we are ultimately led to
postulate an extremely low-entropy state at or near the beginning of the universe. Albert  
calls
this postulate “the past hypothesis”. \cite[15]{Frisch07}    
\end{quote}

It is thus a common view that the familiar time-asymmetric thermodynamic character of our universe, and whatever depends on it, turns on a combination of two things:  (i) a time-\textit{symmetric} probability distribution over the space of possible histories, or `trajectories', allowed by the relevant physical laws; and (ii) a time-\textit{asymmetric} condition, the Past Hypothesis (PH), specifying a low entropy macrostate for the universe shortly after the Big Bang. Conditionalising on PH means disregarding all the trajectories of which it is not true, at the relevant point in time. Compared to our EPR-Bohm toy model, it is like \textit{preselection} rather than \textit{postselection:} the possibilities we ignore are not \textit{actual} but merely \textit{possible.} 

In the urn model, we saw that this difference made a big difference to the counterfactuals. Preselection restricts the range of options we consider in counterfactual reasoning to the preselected class, whereas postselection does not. In an analogous way, many writers have argued that PH is itself the basis of a striking time-asymmetry in counterfactual reasoning. In most contexts, it seems clear that if the present had been different the future would also have been different, but the past would have been the same. It is argued, in effect, that this asymmetry results from the restriction PH imposes on the ways in which the past might be different.

A similar intuitive time-asymmetry is our sense that the past is `fixed' but the future  `open'. Again, PH claims credit. It would take us too far afield to investigate these issues in detail. We will assume for present purposes that these  proposals are substantially correct. More generally, we will assume that the low entropy early universe is the major source of much, if not all, of the observed time-asymmetry of the familiar universe.\footnote{For discussion of such issues  see 
\cite{Price96, Albert02,Frisch07,Price10,PriceWeslake10,Loewer20, Rovelli21}, for example.} 

 On the cosmological scale, PH thus plays the role of the college library, in our little example above. It is the source of the order we encounter in the universe\footnote{Including, of course, ourselves.} --  the mother of all libraries, to adapt a phrase from \cite[118]{Albert02}, or the mother of all preselections.

\begin{figure}[t]
\centering
\includegraphics[width=7cm]{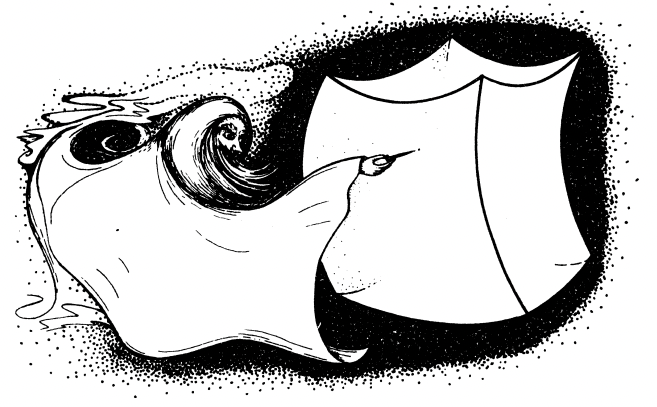}
\caption{Penrose  on the Past Hypothesis 
} \label{fig:god}
\end{figure}

It is difficult to convey how big a restriction on the space of possible trajectories PH needs to be.  \cite[444]{Penrose89}  has a go at it, calculating a figure of $1$ in $10^{{10}^{123}}$, and providing an illustration  of the Creator choosing the initial state with a very, \textsc{very}, VERY fine pin (see Figure~\ref{fig:god}).

Helpfully, Penrose depicts the larger space of possibilities -- the large white box in Figure~\ref{fig:god} -- from with the Creator makes a choice. Let's call this larger space the \textit{Unpinned Regime} (UR). Thus UR is a universe with the same laws as our own, except that it is not governed by PH.\footnote{We needn't take a stand on whether PH should be regarded as a law.} In terms of the standard modern version of the Boltzmann view described earlier, it is what we have on the table if we pause at the first stage: at `(i) a time-symmetric probability distribution
over the space of possible histories \ldots\ allowed by the relevant physical laws', as we put it above.

With UR explicitly on the table, we can now restate our main proposal. It requires us to discuss the statistics of quantum experiments in UR, and some readers may balk at this. Could there be experiments at all in UR? But here we are on well-trodden ground. A community familiar with  Boltzmann Brains should not be troubled by a few simple pieces of quantum optics. 
Such things will arise by chance in a sufficiently expansive UR, and we may discuss their characteristic statistics.\footnote{The measure we need to do so is the one we invoke anyway, at step (i) of the modern Boltzmann program, and also in standard predictions towards the future; see the quote from \cite{Carroll13} in \S\ref{sec:penrose}.}

\section{UR-physics and the open past}\label{sec:ur}

Let us imagine a huge ensemble of instances of the physical structure depicted in Figure~\ref{fig:urn}, happening at random in UR. These instances can be characterised by five-bit spacetime `tiles', just as described in \S\ref{sec:toy}.2. We know that there are subensembles of the big ensemble of such tiles that satisfy EPR correlations. We define such a subensemble by fixing the value of $C$.  We could imagine a kind of Super Observer, outside this universal `URN', picking out such  subensembles by postselection.\footnote{To clarify a point left hanging in \S\ref{sec:toy}.2, notice that UR itself is the world we need, to think of the world as an urn.} 

Just as in our toy model, this process of postselection will not yield counterfactually robust correlations. Concerning a particular random case of Figure~\ref{fig:urn}, the Super Observer has no reason to say that if it \textit{had had} the same A and B settings but a different A outcome, it would also have had a different B outcome. To the extent that such counterfactuals make sense at all, there is no reason to expect this one to hold. As with postselection in our toy model, there's another possibility. The case could have had a different C value, instead. 

In our discussion of the urn model in \S\ref{sec:toy}, we noted two things. First, preselecting the tiles  in the urn (to eliminate, say, $C=1$ cases) does yield counterfactually robust EPR correlations. Second, if we allowed ourselves to think about the history \textit{of a particular tile,} then -- even in the postselected case -- we get a relevant counterfactual. We're inclined to say that if \textit{that particular run} of the experiment had produced a different A outcome, it would nevertheless have had the same $C$ value, which gives us our counterfactual. However, both of these moves are unavailable in UR. There's no preselection, and no basis for holding the past (preferentially) fixed, in thinking about counterfactuals. 

In UR, then, \textit{there are no EPR correlations,} in the usual counterfactually robust sense.\footnote{Or almost none. If UR is sufficiently extended in space and time, it will occasionally produce fluctuations comparable to the region in which we find ourselves. (This was the original Boltzmann-Schuetz hypothesis for explaining the observed low entropy in our universe \cite[Ch.~2]{Price96}.) Within such extremely rare regions, equally likely with either temporal orientation, physics will display the kind of time-asymmetries it does in our own labs, including robust EPR correlations.} Such correlations arise from preselection, permitted and ultimately produced by Penrose's pin. They are products of the Pinned Regime, in other words, and absent from UR. 

Our (provisional) proposal is that this is true of entanglement in general. It is is provisional because, as we note below (\S\ref{sec:general}), there are some cases of (apparent) entanglement where its application is not clear. But we claim that it works for standard cases of EPR and Bell correlations. As we'll see in a moment, the extension to the latter is straightforward.

Readers who are a step ahead of us may have had the following thoughts. First, since there is no time-asymmetry in UR, random EPR experiments in the form of Figure~\ref{fig:urn} should be equally likely with either temporal orientation. Second, since Penrose's pin only applies at one end of the universe, it shouldn't constrain any \upV-shaped instances, in which the vertex points to the future. Are there any examples of such things in our actual universe? If so, do they provide a useful test-bed for the current proposal?

It turns out that the answer to both questions is `yes'. We can find close analogues of the experiment in Figure~\ref{fig:urn} in which the central vertex points in the other direction. And these experiments confirm the proposal, in a striking way. These cases are closely related to some of the most interesting experimental tests of QM's predicted violation of Bell's Inequality. Before we introduce them, it will be helpful to show that our proposal extends easily from EPR correlations to Bell correlations.

\section{From EPR to Bell -- generalising the toy model}\label{sec:tobell}

\begin{figure}[t]
\centering
\begin{tikzpicture}
    \node[draw,rectangle,minimum width=0.6cm,minimum height=0.5cm] (I) at (3.1,-2.5) {\textsf{C}};
\node[] (I0) at (2.5,-3.4) {\textbf{0}};
 \node[] (I1) at (2.9,-3.4) {1};
 \node[] (I2) at (3.3,-3.4) {2};
 \node[] (I3) at (3.7,-3.4) {3};
 \node[] (Prep) at (2.5,-4.7) {\textbf{Preparation}};
 \node[] (C) at (3.1,-4) {$C$};

  \path [dashed,-] (I1) edge (I);
  \path [line width=1.5pt,-] (I0) edge (I);
  \path [dashed,-] (I2) edge (I);
  \path [dashed,-] (I3) edge (I);
  \path [->] (Prep) edge (I0);

    \node[draw,rectangle,minimum width=0.6cm,minimum height=0.5cm] (Abox) at (1.2,0) {\textsf{A}};
   \node[] (Alabel) at (1.2,1.7) {$A$};
    \node[] (A0) at (0.8,1) {0};
    \node[] (A1) at (1.6,1) {1};
    \node[] (a) at (0.5,-1) {$a$};
    \node[draw,rectangle,minimum width=0.6cm,minimum height=0.5cm] (Bbox) at (5,0) {\textsf{B}};
     \node[] (Blabel) at (5,1.7) {$B$};
       \node[] (B0) at (4.6,1) {0};
    \node[] (B1) at (5.4,1) {1};
    \node[] (b) at (5.7,-1) {$b$};

   \path [dashed,-] (Abox) edge (A0);
   \path [dashed,-] (Abox) edge (A1);
   \path [->] (a) edge (Abox.south west);
    \path [dashed,-] (Bbox) edge (B0);
   \path [dashed,-] (Bbox) edge (B1);
    \path [->] (b) edge (Bbox.south east);

   \draw [color=blue,line width=1.5pt,-] (I.north) -- (Abox.south east);
   \draw [color=blue,line width=1.5pt,-] (I.north) -- (Bbox.south west);

\end{tikzpicture}
\caption{Generic V-shaped Bell experiment with preparation in state \textbf{0} at C} \label{fig:V}
\end{figure}
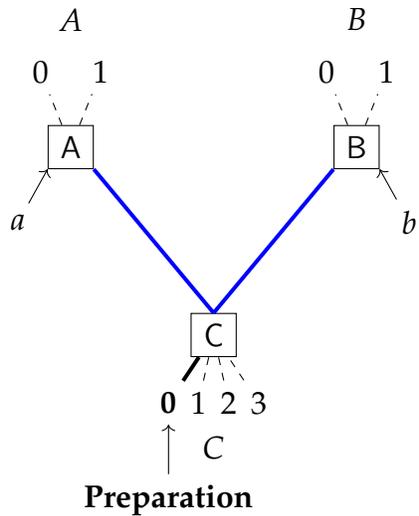

Figure~\ref{fig:V} shows a typical two-particle Bell experiment, with the same V-shaped geometry as the EPR-Bohm experiment of Figure~\ref{fig:urn}. To bring it closer to real-world examples, we have shown four possible initial states at \textsf{C}, which are typically the so-called \textit{Bell states.}   Figure~\ref{fig:V} shows the case in which this is chosen to be state $0$. 

The two entangled particles proceed to measurements at \textsf{A} and \textsf{B}, where the experimeters again choose binary settings $a$ and $b$. In the best experiments of this kind, $a$ and $b$ correspond to different measurement settings on each side. This is done to maximise the predicted violation of the relevant Bell Inequality, which is usually the so-called {Clauser-Horne-Shimony-Holt (CHSH) Inequality.} This means that the experiment has no same-setting cases, which played such a central role in the EPR argument. But there are still two possible settings on each side, so we can code them as binary bits, as before. 

It is easy to imagine a version of this experiment in which the settings at A and B and the initial state at \textsf{C} are all chosen at random (with equal probability for the available options, in all three cases). Since there are now four possibilities at \textsf{C}, we need two binary bits to code them. With this modification, we can record a run of this randomised experiment with a pentagonal tile, just as in Figure~\ref{fig:pent}. The only difference is that $C$ now needs two bits.

If we collect these six-bit tiles  in an urn, as before, and draw them out at random, we find the following close analogues of the EPR-Bohm case. 
\begin{enumerate}\item There are no Bell correlations -- i.e., no correlations requiring a violation of the Bell/CHSH Inequalities -- in the results as a whole. \item We can find Bell correlations by postselecting for any of the subensembles defined by a fixed value of $C$. 
\item These postselected Bell correlations are counterfactually fragile: if we ask whether something at B would have been different, if something at A had been different, the answer will be `No, not necessarily, for the initial state might have been different.' 
\item If we replace postselection with preselection, say by ensuring that all the tiles  in the urn code for $C=0$, we remove this counterfactual fragility; we make it the case that $C$ could not have had a different value, in the relevant counterfactual circumstances. 
\end{enumerate}

All this is exactly as in the original EPR-Bohm urn model, with one important difference. It is much harder to describe the relevant counterfactuals precisely, in the Bell case.\footnote{As noted above (note \ref{fn_1}), one response to Bell's work is to deny that the relevant counterfactuals make sense in QM.} This should not be surprising, if we reflect that sixty years after Bell's initial result, it is still controversial what it implies -- what  the precise kind of dependence (if any!) it requires between the two sides of a Bell experiment. 

We are not going to attempt to do justice to this controversy here.\footnote{Again, see \cite{Myrvold21} for discussion.} Instead, we simply rely on the common name for what Bell is widely held to have discovered: \textit{Bell Nonlocality} (BN). Our point is that preselection from larger, uncorrelated ensembles offers an SOS-style explanation of BN, in two senses. It explains both (i) the existence of Bell correlations in real experiments, and (ii)  their counterfactual robustness  (however exactly that should be characterised). In both respects, the proposal for BN is the same as for EPR correlations.\footnote{In both cases the claim comes with an important qualification. The proposal offers an understanding of EPR/Bell correlations `in terms we can understand', as Penrose put it. It does not explain why the world avails itself of this option, or why it only seems to do so in QM. Those questions take us in the direction of Penrose's second mystery (\S\ref{sec:origins}), which we do not attempt to address here; but see \S\ref{sec:discussion}.4.} 

\section{Does the proposal work the other way up?}\label{sec:does}

If the source of the constraint on the selections responsible for EPR-Bell correlations is, ultimately, Penrose's pin at the beginning of time, then this suggests the following challenge. Suppose we could find a way to perform our EPR-Bohm experiments which put the options not at C in the past, but in the future. There, there is no pin, and so nothing in the world to enforce a selection, apparently.\footnote{In \S\ref{sec:future} we will encounter a proposal for an exception.} If the story is right, that means we shouldn't expect counterfactually robust EPR-Bell correlations, in that case.

Happily, we can do this, and it gives the answer we want. To make it appropriately comparable to our V-shaped EPR-Bohm case, we want to preserve the temporal orientation of the laboratories at A and B. The trick to doing this is to use not a \upV-shaped experiment, flipping the whole V-shaped experiment upside down, but a W-shaped experiment, which just flips the central vertex. Remarkably, we can set things up so that Alice and Bob (the experimenters at A and B, respectively) don't know which version of the experiment they are participating in; more on this below.

Figure~\ref{fig:W} depicts an important experimental protocol, used for some of the best tests of Bell correlations. We call it the W-shaped protocol, referring to the shape of its arrangement in spacetime. It relies on so-called \textit{entanglement swapping.} Two pairs of entangled particles are produced independently, one pair at the source \textsf{S$_1$} and the other at the source \textsf{S$_2$}. One particle from each pair is sent to a measurement device at \textsf{M}, which performs a particular joint measurement on the two particles -- a measurement that has four possible outcomes, shown in Figure~\ref{fig:W} as 0, 1, 2 and 3.

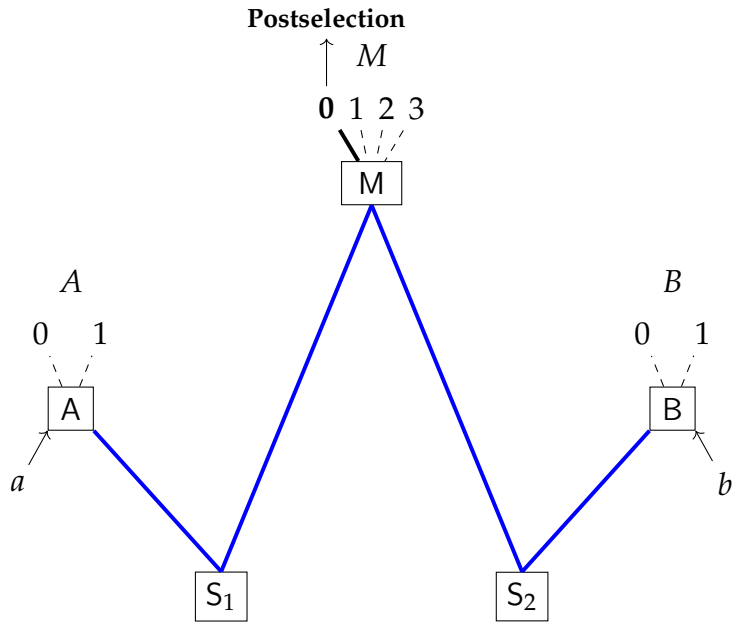
\begin{figure}[t]
\centering
\begin{tikzpicture}
    \node[draw,rectangle,minimum width=0.6cm,minimum height=0.5cm] (I1) at (3,-2.5) {\textsf{S$_1$}};
    \node[draw,rectangle,minimum width=0.6cm,minimum height=0.5cm] (I2) at (7,-2.5) {\textsf{S$_2$}};
    \node[draw,rectangle,minimum width=0.8cm,minimum height=0.5cm] (W) at (5,3) {\textsf{M}};
\node[] (W0) at (4.4,4) {\textbf{0}};
\node[] (W1) at (4.8,4) {1};
 \node[] (W2) at (5.2,4) {2};
 \node[] (W3) at (5.6,4) {3};
\node[] (Out) at (5,4.7) {$M$};

 \path [dashed,-] (W1) edge (W);
 \path [line width=1.5pt,-] (W0) edge (W);
\path [dashed,-] (W2) edge (W);
 \path [dashed,-] (W3) edge (W);

    \node[draw,rectangle,minimum width=0.6cm,minimum height=0.5cm] (Abox) at (1,0) {\textsf{A}};
  \node[] (Alabel) at (1,1.7) {$A$};
    \node[] (A0) at (0.6,1) {0};
    \node[] (A1) at (1.4,1) {1};
    \node[] (a) at (0.3,-1) {$a$};
    \node[draw,rectangle,minimum width=0.6cm,minimum height=0.5cm] (Bbox) at (9,0) {\textsf{B}};
      \node[] (Blabel) at (9,1.7) {$B$};
       \node[] (B0) at (8.6,1) {0};
    \node[] (B1) at (9.4,1) {1};
     \node[] (b) at (9.7,-1) {$b$};
       \node[] (bc) at (4.4,5.2) {\footnotesize{\textbf{Postselection}}};
       \path [<-] (bc) edge (W0);

   \path [dashed,-] (Abox) edge (A0);
   \path [dashed,-] (Abox) edge (A1);
    \path [->] (a) edge (Abox.south west);
    \path [dashed,-] (Bbox) edge (B0);
   \path [dashed,-] (Bbox) edge (B1);
   \path [->] (b) edge (Bbox.south east);
      \draw [color=blue,line width=1.5pt,-] (I1.north) -- (Abox.south east);
   \draw [color=blue,line width=1.5pt,-] (I2.north) -- (Bbox.south west);
      \draw [color=blue,line width=1.5pt,-] (I1.north) -- (W.south);
       \draw [color=blue,line width=1.5pt,-] (I2.north) -- (W.south);

  
\end{tikzpicture}
\caption{The W-shaped case with postselection for outcome $0$ at \textsf{M}} \label{fig:W}
\end{figure}

If we postselect the cases in which the measurement \textsf{M} takes a particular value -- say, $0$, as shown in Figure~\ref{fig:W} -- then each remaining pair of particles exhibits the correlations typical of entanglement, \textit{even though the particles concerned have never interacted}. This is entanglement swapping, and it can be used to test for Bell correlations. Some of the most impressive  results, the so-called `loophole-free' Bell experiments, use this protocol.\footnote{See \cite{Hensen15,Giustina15,Shalm15}.} 

Arrangements for such a Bell experiment are shown on the two wings of Figure~\ref{fig:W}. There, the set-ups are exactly as in the corresponding parts of the V-shaped experiment in Figure~\ref{fig:V}. Each of the measuring devices \textsf{A} and \textsf{B} has two possible settings, chosen by the experimenters at those locations, and two possible outcomes. Each run of the experiment thus produces four binary numbers \textit{a, A, b, B,} comprising two settings and two outcomes. If we add two bits to code for M, each run of the experiment can be described by a six-bit tile,  just as in the V-shaped case.  

Indeed, this W-shaped experiment can be arranged so that the predicted correlations, for each value of \textit{M,} exactly match those of a  V-shaped experiment, for the corresponding value of \textit{C.}  In principle, in fact, we could perform these experiments so that Alice and Bob, at A and B, could not tell whether the experiment had the shape of Figure~\ref{fig:W} or Figure~\ref{fig:V}. And if we use the W-shaped experiment to generate six-bit tiles  for our urn model, there is nothing in the statistics of the urn to tell us whether it was produced by the apparatus in Figure~\ref{fig:V} or Figure~\ref{fig:W} (or some combination of the two). 

These W-shaped experiments can be performed with the measurement \textsf{M} in any one of three different temporal locations, with respect to those at \textsf{A} and \textsf{B}. \textsf{M} can be in the past (i.e., in the overlap of the past light cones of \textsf{A} and \textsf{B}), in the future (i.e., the overlap of the future light cones), or spacelike separated from \textsf{A} and \textsf{B}. (There are also some mixed cases, which we can ignore here.) In the three  experiments cited above, two used the version in which \textsf{M} was in the past with respect to \textsf{A} and \textsf{B}, and one used spacelike separation. 

The case that interests us, and the one actually depicted in Figure~\ref{fig:W}, is the remaining possibility, in which \textsf{M} occurs strictly \textit{later} than \textsf{A} and \textsf{B}. This case relies on \textit{delayed-choice} entanglement swapping (DCES). So far as we are aware, no loophole-free Bell experiment has been performed in this version, although the experimental predictions are the same as in the other two cases (and other kinds of delayed-choice entanglement swapping measurements have been performed \cite{Ma12}, with successful results). 

Several authors have argued that in the delayed-choice version of the W-shaped experiment, Bell correlations in the postselected ensemble should be regarded as selection artefacts, a manifestation of collider bias.\footnote{See \cite{Gaasbeek10,Egg13,Fankhauser19,Guido21,PriceWharton21a, Mjelva24}.} In orthodox QM, the argument is simple. Because the measurements at A and B occur \textit{before} the measurement at M, the orthodox account goes like this. The measurement at A collapses the state of the inner particle of the same pair into a state that reflects both the setting \textit{a} and the outcome \textit{A}. In effect, the inner particle then `knows' what measurement has been performed at A, and what its result was. (This is simply the EPR correlations at work.) The same is true for the measurement at B, and the upshot is that the measurement at M is performed on two particles that `know' what happened at A and B, respectively. Bracketing concerns about the orthodox model itself, the causal pathways to the collider look straightforward. If we are happy to treat collapse as a physical process, it is natural to regard this as a mechanistic model. 

As has been pointed out, this means that the argument for the existence of Bell Nonlocality between A and B in these cases is unreliable.\footnote{In \cite{PriceWharton21a} we call this the \textit{Collider Loophole} for W-shaped tests of Bell Nonlocality, and discuss which W-shaped Bell tests are subject to it. Here we need only the DCES case, where it is uncontroversial.} Because a change at A could produce a change at M, there is no need for it to produce a change at B. Once again, we have chosen to word this vaguely, to avoid tying our hands on the controversial issue of what precise kind of dependence (if any) is involved in BN.\footnote{If we revert to an EPR version of the DCES W experiment, we get the same clear-cut counterfactuals as in the V-shaped EPR case. The same-setting cases show perfect correlation or anticorrelation between A and B, depending on the state at M. On what grounds could one insist that differences at A and/or B could not make a difference at M?\label{fn_36}} The point we need is simply that in the orthodox model of QM, the case for scepticism about BN is clear-cut, in these DCES W experiments. 

The timing of the measurement at M with respect to A and B makes a big difference to this argument. If M occurs before A and B, orthodox QM doesn't give us a collider at M. On the contrary, causal arrows now point \textit{away from} M, towards A and B. The states of the particles there depend on the outcome at M, rather than the other way round. And in this case, there is a strong intuitive reason for retaining the counterfactuals on which BN between A and B depends. We are inclined to say that a difference at A could not make a difference at M, because M is in the past. We might find this difference puzzling, given that the correlations in the W-shaped case are insensitive to the location of M, but for the moment it simply illustrates one of the intuitive time-asymmetries of the Pinned Regime.\footnote{For further discussion, see \cite{PriceWharton21b}.}

Returning to the DCES case, we have seen that orthodox QM delivers a verdict of collider bias. What about our present framework? We don't have the causal arrow provided by collapse, but instead we have a claim about counterfactual dependence between the two particles in each of the wings of the W, grounded on ConCorrFs at \textsf{S$_1$} and \textsf{S$_2$}. But this story is too underdeveloped, and too operational, for  application to this new case. As we noted, the orthodox model might well be regarded as proposing a mechanism, whereas our stance throughout has been SOS, not MS. 

It would be pleasing to have an alternative ontological model to orthodox QM, and be able to show in terms of it that the inputs to M are counterfactually dependent on conditions at A and B. If our main proposal is to succeed, this dependence would need to be associated with the claim that we have ConCorrFs at \textsf{S$_1$} and  \textsf{S$_2$}, in the two wings of the experiment. At present, however, our proposal is purely operational, and hence not suited for this kind of mechanistic,   `under the hood' work.

For the moment, then, we need to fall back on SOS, and the lessons of our urn models. We have noted that Figure~\ref{fig:V}  and  Figure~\ref{fig:W} generate the same six-bit urn models. In both cases we need to randomise the settings at A and B. In Figure~\ref{fig:V} we also need to randomise the inputs at C. But in Figure~\ref{fig:W} the world does the randomising of the variable at M, which is now an output. That's a useful point, to which we'll return below. 

With this small difference, we generate the same urns of six-bit tiles  in both cases. We know that postselection from these urns can give us Bell correlations but not counterfactual robustness, in either case. What does generate counterfactual robustness, in the case of Figure~\ref{fig:V}? Applying the lessons of our original EPR urn model, it is one of two things: either preselection, or -- looking `through' the urn to the actual run of the experiment that generates a particular tile -- some `hold the past fixed' principle of counterfactual reasoning. Neither of these options is available in the DCES W-based urn model. We can't preselect the results at M, at we did at C, and there is no `hold the future fixed' principle to constrain the result there, in the actual experiment. 

The DCES W case thus confirms the conclusions we drew about the V case in UR. Because Penrose's pin constrains the past but not the future, the future in our actual (pinned-in-the-past) universe is comparable to the past in UR. This means that the future-pointing vertex of the DCES W case is comparable to a V in UR. Lacking anything to constrain the CorrF at that point, there is no foundation for the counterfactuals required for BN 
 between A and B. All we have is postselection, which is counterfactually fragile.\footnote{Again, we could revert to an EPR version of the DCES W experiment, to make the point with clear-cut counterfactuals; cf.~note \ref{fn_36}.} 

There is another difference between the W and V cases, which works to our advantage. In generating the V-based urn models, we put the initial distribution over the initial states in by hand. In the W-based case, as we just noted, there is no need to do that. The raw QM probabilities gave us the flat distribution we needed. By time-symmetry, this suggests that the same will be true of the V cases in UR, answering the challenge we raised in \S\ref{sec:application} about the source of the measure.

We can take the discussion of the W case one step further. The argument above suggests that if there were something in physics to restrict the outcome of the measurement at M to a single value, we would have a ConCorrF at that point, as we do in V-shaped cases in the ordinary world. This would restore the counterfactuals needed for BN. It turns out that we can find such a proposal in the literature.

\section{Constraint from the future?}\label{sec:future}

In DCES W Bell tests, then, Bell correlations are regular, counterfactually-fragile selection artefacts, rather than signs of genuine, counterfactual-supporting BN. In our terminology, the central measurement at M is a CorrF, but not a \textit{Constrained} CorrF. In the V-shaped cases, by contrast, \textit{constraint} at the central vertex comes from Initial Control, and ultimately from PH -- from Penrose's pin. There is nothing like that in the future, in normal cases, to constrain the central vertex of a DCES W case.

 \begin{figure}[t]
\centering
\begin{tikzpicture}
 \node[draw,rectangle,minimum width=0.6cm,minimum height=0.5cm] (I1) at (3,-2.5) {\textsf{S$_1$}};
    \node[draw,rectangle,minimum width=0.6cm,minimum height=0.5cm] (I2) at (7,-2.5) {\textsf{S$_2$}};
    \node[draw,rectangle,minimum width=0.8cm,minimum height=0.5cm] (W) at (5,3) {\textsf{M}};
\node[] (W0) at (4.4,4) {\textbf{0}};
\node[] (W1) at (4.8,4) {1};
 \node[] (W2) at (5.2,4) {2};
 \node[] (W3) at (5.6,4) {3};
  \node[] (Out) at (5,4.6) {$M$};
   \node[] (bc) at (4.4,5.2) {\footnotesize{\textbf{Boundary constraint}}};


 \path [dashed,-] (W1) edge (W);
 \path [line width=1.5pt,-] (W0) edge (W);
\path [dashed,-] (W2) edge (W);
 \path [dashed,-] (W3) edge (W);
\path [->] (bc) edge (W0);

    \node[draw,rectangle,minimum width=0.6cm,minimum height=0.5cm] (Abox) at (1,0) {\textsf{A}};
  \node[] (Alabel) at (1,1.7) {$A$};
    \node[] (A0) at (0.6,1) {0};
    \node[] (A1) at (1.4,1) {1};
    \node[] (a) at (0.3,-1) {$a$};
    \node[draw,rectangle,minimum width=0.6cm,minimum height=0.5cm] (Bbox) at (9,0) {\textsf{B}};
      \node[] (Blabel) at (9,1.7) {$B$};
       \node[] (B0) at (8.6,1) {0};
    \node[] (B1) at (9.4,1) {1};
     \node[] (b) at (9.7,-1) {$b$};

   \path [dashed,-] (Abox) edge (A0);
   \path [dashed,-] (Abox) edge (A1);
    \path [->] (a) edge (Abox.south west);
    \path [dashed,-] (Bbox) edge (B0);
   \path [dashed,-] (Bbox) edge (B1);
   \path [->] (b) edge (Bbox.south east);

      \draw [color=blue,line width=1.5pt,-] (I1.north) -- (Abox.south east);
   \draw [color=blue,line width=1.5pt,-] (I2.north) -- (Bbox.south west);
      \draw [color=blue,line width=1.5pt,-] (I1.north) -- (W.south);
       \draw [color=blue,line width=1.5pt,-] (I2.north) -- (W.south);


\end{tikzpicture}
\caption{W case with outcome $0$ imposed by a boundary constraint}\label{fig:W2}
\end{figure}
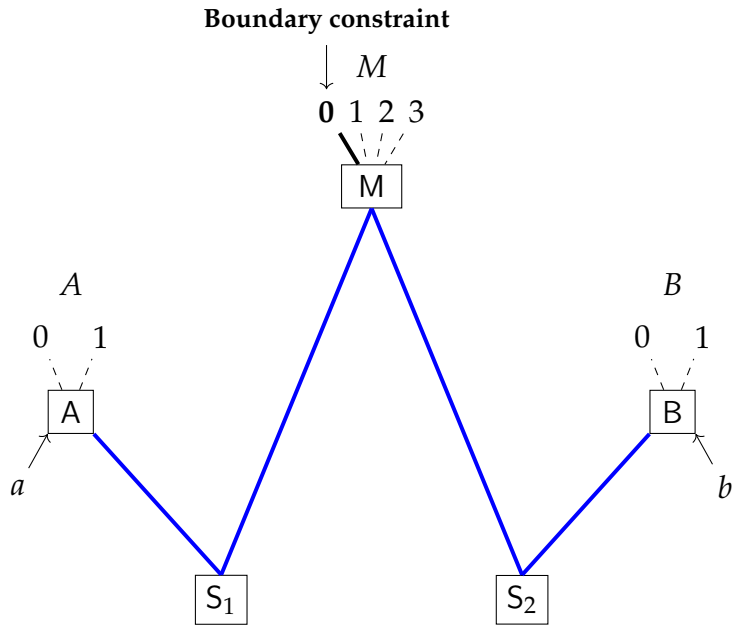

However, there is an exceptional case  in the physics literature, which adds a constraint to a DCES W case in the way depicted in Figure~\ref{fig:W2}.
It is a proposal from Horowitz and Maldacena about the   black hole information paradox \cite{HorowitzMaldacena04}. 
The background is that Stephen Hawking discovered a process (now called `Hawking radiation'), by which black holes eventually evaporate. He thought initially that this process would be random, preventing the escape of information falling into the black hole in the past, and hence in conflict with the usual reversibility, or `unitarity', of quantum theory. Various mechanisms to allow escape of this information were then proposed, including the one that interests us here.

Horowitz and Maldacena describe their proposal as follows:  
\begin{quote}
In [Hawking's] process of black hole evaporation, particles are created in correlated pairs with one
falling into the black hole and the other radiated to infinity. The correlations remain even
when the particles are widely separated. The final state boundary condition at the black
hole singularity acts like a measurement that collapses the state into one associated with
the infalling matter. This transfers the information to the outgoing Hawking radiation in
a process similar to ``quantum teleportation''. \cite{HorowitzMaldacena04}
\end{quote}
In effect, Maldacena and Horowitz propose that a measurement at the black hole singularity creates a zigzag causal path, along which information can escape from a black hole. It is crucial that this process supports counterfactuals. If it is to explain how information escapes from black holes, then it needs to be counterfactually robust: if different information had fallen into the black hole in the first place, different information would have emerged. And the trick needed to ensure this is that unlike ordinary entanglement-swapping measurements, this one has only one possible value. In our terms,  it is a \textit{Constrained} CorrF.

Discussing the Horowitz-Maldacena hypothesis recently, Malcolm Perry puts it like this:

\begin{quote}
The interior of the black hole is therefore a strange place where
one's classical notions of causality \ldots{} are violated. This does
not matter as long as outside the black hole such pathologies do not
bother us. \cite[9]{Perry21}
\end{quote}
Our proposal is that such pathologies are actually common outside black holes, where, under the name entanglement, they  have been bothering us for decades. The  boundary conditions responsible are exceedingly common, being just the familiar, controllable constraints on the initial conditions of experiments.\footnote{They don't need black holes, but they may ultimately depend on the physics of a singularity at the other end of the universe.}

For the moment, the Horowitz-Maldacena hypothesis is further confirmation that the comparison between the V-shaped and DCES W-shaped experiments gives the answers our proposal requires. In our ordinary world, with a pin in the past but not the future, the central vertices are constrained in the V-shaped cases but not in the DCES W-shaped cases. The V cases lose their constraint in UR, while the W cases acquire a constraint in a Horowitz-Maldacena black hole. All this is as it should be, if the origin of familiar EPR-Bell counterfactuals is Penrose's pin, as we propose.

 \section{Does the proposal generalize?}\label{sec:general}

We have proposed an SOS account of Bell correlations in some important cases, but does it generalise? Dealing with some canonical entanglement experiments is a reason for optimism, but can we do better? Alternatively, can we show that it does \textit{not} generalize to some entangled systems? 

The important question seems to be this. Given an arbitrary entangled system S, can we always find a natural `super-ensemble' \textbf{S} in our UR, containing S as one of its members, with the following properties?

\begin{enumerate}[label=(\roman*)]
    \item In the natural time-symmetric measure on UR, \textbf{S} as a whole washes out the correlations taken to be manifestations of  entanglement in S. \item We can restore these original correlations by imposing the actual initial state of S as a boundary condition. \end{enumerate} 

For cases in which the entangled state in question has an initial preparation, as in the V-shaped examples, this seems to be straightforward. In QM it is standard to use a `density matrix', $\rho$, to represent states with an uncertain preparation.  In this framework, it is easy to show that a completely uncertain preparation (over a sufficiently large number of possibilities) will always result in what is known as a `maximally mixed state'.  Such states have exactly the properties required here; there are no correlations which can be observed in a maximally-mixed state.  Yet, if one selects one particular preparation state $|\psi\rangle$, the density matrix must take the form $\rho=|\psi\rangle\langle\psi |$, which can never be maximally mixed.  So quantum correlations can always be restored by exercising initial control on a prepared quantum system. 

Conversely, removing this initial control -- replacing it with an equal-probability set of preparations which span the Hilbert space -- always removes these correlations.  This makes the initial control entirely responsible for the eventual correlations, for any jointly-prepared entangled state. 

However, there are further ways to generate entangled states, different from both the V- and W-shaped examples. Examples include the Hong-Ou-Mandel effect \cite{Hong87}, and so-called `which-way' entanglement \cite{Ferrari10}.\footnote{Though the latter cases, at least, can always be reduced to more conventional entanglement scenario, where each path can be recast as an entangled state of zero-particle and one-particle modes \cite{catani23}.}   As it stands, the present proposal does not capture these cases.

 \section{Discussion}\label{sec:discussion}
\subsection{Reversing the argument}
We have argued that EPR and Bell correlations can be understood as selection artefacts. In summarising the argument, it is helpful to reverse the order of presentation,  beginning with the cases in which the proposal is least contentious. Reframed in this way, the argument has two main steps. 

\textbf{First,} we note that for \textit{some} EPR-Bell experiments -- the DCES W cases -- it is uncontroversial that the correlations they produce are selection artefacts. The key points are as follows.\begin{enumerate}
   
    \item In orthodox QM these cases are straightforward collider bias, in a causal model provided by the collapse ontology. Bracketing concerns about the orthodox ontology, this is an MS-level explanation of Bell correlations as selection artefacts, in these cases. 
    
    \item In our SOS picture, all the key elements are also straightforward: 
    \begin{itemize}
    \item The experiment itself provides the larger uncorrelated ensemble, with four possible measurement outcomes at M.
    \item The means of postselection from this larger ensemble is familiar, widely used elsewhere in QM (e.g., in quantum teleportation).
    \item Because it is postselection, the resulting correlations are counterfactually fragile. 
\end{itemize}
\item The central measurement of a DCES W case is a CorrF, in our terminology, but not yet a ConCorrF. The Horowitz-Maldacena hypothesis describes an exceptional case, in which such a measurement becomes a ConCorrF, supporting robust counterfactuals.

\end{enumerate}
 We take all of this to be uncontroversial, for these W cases. The Horowitz-Maldacena hypothesis is not uncontroversially true, of course, but we don't depend on that. What matters is what would obtain \textit{if it were true,} and that does seem uncontentious.
 
\textbf{Second,} we propose that all of this applies equally in the ordinary two-particle V-shaped cases, with one big difference. These are the key points.
\begin{enumerate}
   \setcounter{enumi}{3}
    \item Speaking in favour of the parallels, we noted that V and W experiments can be constructed with exactly the same correlations, once C replaces M, or vice versa. 
     The big difference turns on what follows from this replacement, in normal circumstances. 
   
    \item C is normally associated with \textit{inputs,} under the control of an experimenter (or of some alternative physical system, itself ultimately `controlled' by Penrose's pin).  M on the other hand is normally associated with \textit{outputs,} uncontrolled in the future in any corresponding way. 
    \item This means that the CorrF at C is normally a \textit{ConCorrF,} constrained by ordinary initial control. It would be a plain CorrF in UR, but that is not our ordinary world.\footnote{It is an interesting question whether there might be real experiments in which the CorrF is unconstrained, e.g., using a QM device to select and record the initial state, delaying observation until the A and B measurements are concluded. We owe this suggestion to Gerard Milburn.} 
    \item For M it is the other way around. The unconstrained case is the normal one, the ConCorrF case being confined (at least within the limits of our present discussion) to Horowitz-Maldacena black holes. 

\end{enumerate}
These points explain how we reconcile two claims, superficially in tension. On the one hand, the V and W cases rely on the same underlying correlations, so that the available ensembles are isomorphic. On the other hand the derived experimental correlations have a very different character  -- counterfactually fragile in one case, counterfactually robust in the other. The difference turns on the different methods of selection normally available to us in the two cases: mere postselection in the DCES W experiments, and preselection enabled by Initial Control in the V experiments. 

\subsection{Is the proposal too simple?}

As we noted in \S\ref{sec:epr-bohm} (note \ref{fn5}), it might be felt that our proposal is \textit{too} simple -- `almost trivial', as one commentator put it. Can it really tell us anything interesting about entanglement? The comparison between the DCES W and the V cases answers this challenge. In the DCES W cases, \textit{there are no robust EPR correlations, or BN, across the W as a whole.} The correlations exist, but they are \textit{mere} selection artefacts, in the everyday, non-counterfactual-supporting sense. This changes when the central measurement at M (in our notation) is \textit{constrained,} as in the Horowitz-Maldacena proposal.\footnote{We noted that if it didn't change,  the 
proposal could not address the blackhole information paradox, because  it wouldn't support the necessary counterfactuals.}

In the DCES W case, then, our simple proposal does some very substantial work. In what has been proposed as a real case, it shows us how robust spacelike EPR and Bell correlations can be a product of two things: a CorrF of the kind we find at M, and the boundary condition, or \textit{constraint,} that turns that CorrF into a ConCorrF. 

The V case is the same, except that we don't need a black hole. Instead, we get the constraint for our ConCorrF at C (in our notation) from ordinary Initial Control of the initial state -- itself ultimately enabled, apparently, by Penrose's pin. The payoff is identical, and equally nontrivial: we get an explanation of the puzzling counterfactual-supporting dependencies between widely-separated systems, exemplified by the EPR and Bell correlations. 

Does this answer the question that Penrose actually had in mind, for his first mystery? We can't be certain, obviously, but it seems a good fit. In any case, with the nature of the payoff clarified, we don't need to lean on Penrose's framing. But as we'll see in a moment,  Penrose's formulation remains helpful in another way. His second mystery helps to clarify what our proposal does \textit{not} explain.

Before we leave the positive side of the story, we want to call attention to a consequence of the simplicity of the proposal. The facts on which it relies are `in the public domain', in an important sense. They are operational correlations, widely agreed to exist in a broad range of approaches to the interpretation of QM. 
As long as we are realists about these \textit{operational} matters,\footnote{A view that rejects such realism is the Algorithmic Idealism of \cite{Muller20}.} and refrain from complicating life with Many Worlds, then we have what the proposal needs: a domain of operational facts with the required structure, from which robust EPR and Bell correlations emerge by selection, in the way we have described. 

This means that within these broad limits, it is hard to see how there could be space for any \textit{other} answer to Penrose's first mystery (interpreted in this way). If we had another, we would have two different answers to the same question, at least one of them relying on facts in the public domain. That would be a surprising result -- unless, perhaps, the two answers actually responded to different questions, in some subtle way. 

One way in which this might happen turns on the distinction we drew in \S\ref{sec:toy}.3 between two explanatory stances, the Structural-Operational Stance (SOS) and the Mechanistic Stance (MS). Returning to that distinction, we want to comment on our terminology for SOS, and then make some remarks about the prospects for MS, in the light of our main proposal.

\subsection{Structuralism}
Why is our proposal appropriately called `Structural-Operational'?
Taking `structural' first, there is a use of this term in philosophy of physics that derives, amongst other sources, from the work of R.I.G.~Hughes \cite{Hughes89,Hughes93}.\footnote{See \cite{Felline10} for a more recent survey; we are grateful to Steven French here.}   
Rob Clifton \cite{Clifton98} borrows the following gloss from \cite[132]{Hughes93}.
\begin{quote}
We explain some feature \textit{B} of the physical world by displaying a
mathematical model of part of the world and demonstrating that
there is a feature \textit{A} of the model that corresponds to \textit{B,} and is
not explicit in the definition of the model. \cite[7]{Clifton98}  
\end{quote}
Clifton continues:

\begin{quote}
    It is natural to call explanations based on this maxim \textit{structural} to emphasize
that they need not be underpinned by causal stories and may make essential reference to purely mathematical structures that display the similarities
and connections between phenomena. \cite[7]{Clifton98}
\end{quote}
Later he emphasises the compatibility, as he sees it, between structural and causal stances.
\begin{quote}
     I should also emphasize that I do not take structural
explanation to be incompatible with explanation by appeal to causal structures. My claim is only that explanation \textit{as} explanation does not privilege
one sort of structure over any other. 
\cite[19]{Clifton98}
\end{quote}
This message of compatibility will be useful to us in a moment, but first, in the light of these sources, why is the term Structural-Operational a good fit for our proposal? 

Three points seem relevant: (i) the proposal appeals to a simple mathematical model of the operational correlations in the various V and W experiments we have discussed; (ii) counterfactually-robust EPR and Bell correlations are, in Hughes's words, `not explicit in the definition of [this] model'; but (iii) such correlations are readily found within it, once we have the operation of constraining the initial state at C (or final state at M, in the constrained DCES W case). Thus the present proposal is \textit{structural} in Hughes's sense, and \textit{operational} in an obvious sense.

\subsection{Mechanisms again?}

Can we clarify what our proposal does \textit{not} explain? We can approach this issue via 
 a natural challenge. As we saw in \S\ref{sec:toy}.3, the same SOS-level proposal could be applied to classical cases, such as our simple device to generate EPR correlations. Indeed, as we noted in \S\ref{sec:time-asymm}, the entire modern Boltzmannian  explanation of the time-asymmetry of our universe might be regarded as an application of the same methodology. It generates plenty of correlations -- think of outgoing concentric ripples on a pond, for example\footnote{Such obvious classical examples involve the coordination of large numbers of components, and hence seem to depend on sources of low entropy. Are there classical cases that are not like this? Either way, the answer seems likely to be helpful in considering what distinguishes the classical and quantum cases.} -- but they are not quantum in nature. Hence the challenge: even if our proposal does something to explain the peculiar nonlocal character of entanglement, it doesn't put its finger on the distinctively quantum element. 

This points us in the direction of Penrose's second mystery:
\begin{quote}
[W]hy is [entanglement] something that we barely notice in our direct experience of the world? Why do these ubiquitous  effects of entanglement not confront us at every turn?\end{quote} 
Reframing these questions a little, we could put the challenge like this. If our proposal is correct, then why does the simple structure it describes seem to be confined to the quantum world? This isn't the only question our hypothesis doesn't answer, of course,\footnote{Another obvious candidate concerns the strength of the quantum correlations. Why Bell correlations, and not something stronger? See \cite{Bub13} for discussion. And why not signalling? See \cite{Muller20}.} but it seems a particularly pressing one, given the simplicity of the proposal.
 
It seems plausible that the answer is related to difficulties of classical mechanisms in the QM case. If so, then we might try to address it by returning to MS, and asking whether the proposal of this paper is helpful in thinking about distinctively \textit{quantum} mechanisms. There is one obvious hook. In \S\ref{sec:does} we noted that in orthodox QM, with collapse treated as a real physical process, we have a mechanism for a collider at the central vertex of a DCES W experiment. In looking for alternatives, presumably, we want the same thing: some quantum ontology to serve as inputs to the measurement at that vertex, itself influenced by the settings and outcomes of the measurements at A and B.

From this point, the respect for time-symmetry embodied in our use of the W/V comparison counts in favour of a similar story at the central vertex of a V experiment. What comes out of the V (at least in UR) should be comparable to what goes into the central vertex of the DCES W -- and correlated with the settings and outcomes of measurements at A and B, now later in time, in just the same way. 

This suggests that the natural MS complement to the present proposal will be a retrocausal model, in some sense \cite{PriceWharton15, FriedrichEvans19, WhartonArgaman20}. However, such models usually focus on {\textit{hidden variables}} (traditionally denoted by $\lambda$), taken to be influenced  by future measurement {\textit{settings.}} The present framework may diverge  in two ways.\begin{enumerate}
    \item From the SOS point of view, the target variable is the input state at C (by analogy with the output state at M), rather than some hidden variable  $\lambda$. 
    \item We may need to think of the relevant influences at A and B as being both the settings and the outcomes, rather than just the settings. In the orthodox QM treatment of the DCES W case, after all, the measurement at M is influenced by both pieces of information, from both sides.
    \end{enumerate}

 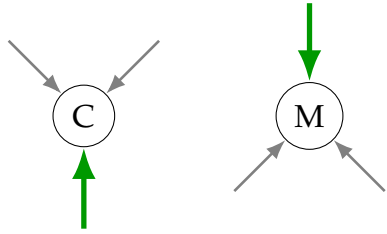
\begin{figure}[t]
\centering
\begin{tikzpicture}
    \node[circle,draw] (C) at (1,1.5) {{C}};
\coordinate (1c) at (1,0);
\coordinate (2c) at (0,2.5);
\coordinate (3c) at (2,2.5);
  \path[color=green!60!black,rounded corners,line width=2pt] (1c) edge (C);
   \path[color=gray,rounded corners,line width=1pt] (2c) edge (C);
     \path[color=gray,rounded corners,line width=1pt] (3c) edge (C);

         \node[circle,draw] (M) at (4,1.5) {{M}};
\coordinate (1m) at (4,3);
\coordinate (2m) at (3,0.5);
\coordinate (3m) at (5,0.5);
  \path[color=green!60!black,rounded corners,line width=2pt] (1m) edge (M);
   \path[color=gray,rounded corners,line width=1pt] (2m) edge (M);
     \path[color=gray,rounded corners,line width=1pt] (3m) edge (M);

\end{tikzpicture}
\caption{Three-way control, V and W cases} \label{fig:tri}
\end{figure}

\noindent In exploring connections between conventional retrocausal models and the present proposal, it may be helpful to think of the variable at C or M as being influenced in three directions: two converging arrows to the central vertex of the V or W, and a third arrow from outside, in the case in which the state at that vertex is constrained (see Figure~\ref{fig:tri}). After all, at the heart of the proposal is the claim that we need to pay careful attention to the constraints shown in green.

This brings us to the final topic we want to put on the table: the implications of the proposal for causal modeling in general. If we are correct in proposing that these kinds of constraints play a crucial role in explaining entanglement, they should certainly be on the radar in that field.

\subsection{Implications for causal modeling}

We noted earlier that our proposal involves two novel features, by the lights of standard causal modeling. One is the kind of constraint just mentioned, the other the existence of Correlating Forks (CorrFs) with the `open to the future' temporal orientation normally associated with common causes (as in Figure~\ref{fig:cc}). Familiar cases of CorrFs are common effects, or colliders, with the orientation of 
Figure~\ref{fig:ce}.

We avoided the term `collider bias' above, so as not to be forced into the use of causal models where they were unnecessary and potentially confusing. We relied instead on the underlying statistical phenomena of Simpson's Paradox and Berkson's bias. But let's now re-embrace the familiar causal terminology, in order to explain the notion of constraint.

As we said in \S\ref{sec:simpson}, a collider is a vertex in a DAG at which two (or more) causal arrows converge. It will be convenient to use an example
 already well-known in decision theory and the philosophy of  causation, the so-called Death in Damascus case \cite{Gibbard78}. Suppose that you and Death are each deciding where to travel tomorrow (Figure~\ref{fig:M1}). You have the same two possible destinations -- in the usual version, Damascus and Aleppo. Your choice and Death's choice both influence the (aptly named) collider variable, which determines your survival. Let this variable take value $0$ if you and Death do not meet, and $1$ if you do. (We assume for simplicity that if the two of you choose the same destination, you will meet; and that this will be fatal, from your perspective.)

\begin{figure}[t]
\centering
\begin{tikzpicture}
    \node (a) at (0,0) {Your choice};
    \node (F) at (3,3.3) {A meeting?};
    \node (b) at (6,0) {Death's choice};
\coordinate (coll) at (3,3); 
  \path[color=gray,rounded corners,line width=2pt] (a) edge (coll);
   \path[color=gray,rounded corners,line width=2pt] (b) edge (coll);

\end{tikzpicture}
\caption{A collider (`Death in Damascus')} \label{fig:M1}
\end{figure}
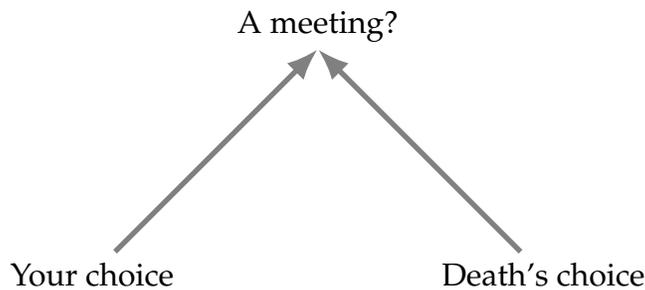

The example illustrates some points we outlined in \S\ref{sec:simpson}. 
If we sample statistics for this case, across a large population, they may suggest that people have an uncanny ability to evade Death, always choosing the opposite destination. If so, it is because we have introduced selection bias, by interviewing survivors only; such cases are sometimes called `survivorship bias' \cite{Czeisler21}.  Survivor bias is a species of collider bias, or Berkson's bias. The collider is a CorrF, in the terminology we introduced earlier: conditioning on the collider  induces a correlation between your choice and Death's choice.

If you are a survivor, you may think to yourself, `I'm a survivor, so if I had chosen the other destination, Death would also have made a different choice.' You would be wrong. If you had chosen the other destination, you wouldn't have been a survivor. This illustrates the fact that correlations resulting from conditioning on a collider do not support counterfactuals, in normal circumstances.

Colliders are very familiar in causal modeling, but our proposal requires an unusual modification, what we have elsewhere called a \textit{constrained} collider \cite{PriceWharton21b,PriceWharton22}. Constraint here is a restriction imposed from outside a causal model, biasing or completely specifying the value of the collider variable in question.

Thus in Death in Damascus, your choice and Death's both influence the variable that takes value $1$ if you meet, and $0$ otherwise. If Fate wants to ensure that your number's up, as it were, she \textit{constrains} this collider variable, setting its value to $1$. In effect, Fate imposes a future boundary condition, \textit{requiring} that the collider variable take value $1$.

This boundary condition makes a big difference to the counterfactuals. If it weren't for Fate's role, your grieving relatives would be entitled to say, `If only you had made the other choice, you would still be with us today!' Once Fate constrains the collider, this is no longer true. If you'd made the other choice you would have met Death in the other place, instead. 

With Fate constraining the collider, then, there is a counterfactual-supporting connection between your movement and Death's.  You control Death's movements, in effect, as \cite{PriceWeslake10} point out.\footnote{Price and Weslake are discussing the proposal that the ordinary direction of causation rests on the Past Hypothesis \cite{PriceWeslake10}. They use the Death in Damascus case to point out that a seemingly analogous future boundary constraint might produce not future-to-past causation, but the kind of zigzag influence just described. They suggest that a boundary constraint in the past might be expected to produce zigzags via the past, but don't connect this suggestion to quantum entanglement, as we have done above.}  In \cite{PriceWharton21b,PriceWharton22}  we called this \textit{Connection across a Constrained Collider} ({CCC}) -- the terminology was deliberately non-committal about whether we regard it as causation, strictly speaking.

If this kind of constraint were confined to fictional cases, we might doubt its relevance for causal modelers. But QM provides real cases. For a start, it gives us the DCES W case (Figure~\ref{fig:W}), where in orthodox QM, as we noted in \S\ref{sec:does}, the outcome of the measurement M is certainly a collider. Conditioning on that collider induces correlations -- including EPR or Bell correlations, depending on the precise experiment we have in mind -- between the factors at A and B that influence the outcome at M. As we saw, the Horowitz-Maladacena hypothesis is that this collider variable is \textit{constrained} (in our terminology) when the measurement takes place at the final singularity of a black hole (Figure~\ref{fig:W2}). This is  a \textit{postulated} real case of collider constraint; if you want to do causal modeling on the Horowitz-Maladacena proposal, you need this notion. 

Still, the most interesting applications of the notion of constraint are in the past, not the future. To bring them into view, it is helpful to revert to statistical level of description, talking again of Berkson's bias, or simply selection bias,  rather than specifically of collider bias. We noted that the DCES W experiment (Figure~\ref{fig:W} and the V experiment (Figure~\ref{fig:V}) can be set up to display identical correlations, once M and C are mapped into each other in the obvious way. This implies that the results of conditioning on the input at C are the same as conditioning on the output at M -- we have exactly the same equations, with the $C$ value substituted for the $M$ value. Conditioning on $C$ induces just the same correlations between A and B in Figure~\ref{fig:V} as conditioning on $M$ does in Figure~\ref{fig:W}. That's why Figure~\ref{fig:V} is a CorrF, in our terminology. If we wanted a specific name for the resulting cases of Berkson's bias, then, as in \S\ref{sec:penrose}, we could call them {predecessor bias.} 

Thus CorrFs with this novel `open to the future' temporal orientation are not merely real, but exceedingly common. This is the second interesting consequence of our discussion for causal modeling, and it doesn't depend on the correctness of our main proposal. As we have just seen, the existence of such CorrFs falls directly out of the operational correlations of common entanglement experiments. Moreover, while it may have been helpful to get there via the DCES W case, where the language of colliders is easily invoked, the point we need was already present in the operational  correlations in the V case, as we used them in \S\ref{sec:toy}.

Once we have identified the CorrF in the V case (Figure~\ref{fig:V}), the notion of constraint finds its most striking application. Unlike in the DCES W case, typical V-shaped CorrFs of this kind are \textit{constrained,} in our sense. Ordinary preparation of the initial state of the experiment does the job. These ubiquitous `open to the future' Conjunctive Forks are typically \textit{Constrained} Conjunctive Forks (ConCorrFs). We have proposed that this constraint is the origin of the counterfactual robustness of EPR and Bell correlations. We recommend the proposal to the causal modeling community, as well as to readers in quantum foundations. 

\subsection*{Acknowledgments}
We are indebted to many people for helpful comments on previous versions of this material. They include: Emily Adlam, Holly Andersen, Nicolas Berggruen, Časlav Brukner, Jeff Bub,  Eric Cavalcanti, Giulio Chiribella, Michael Cuffaro, George Davey Smith, Steven French, Simon Friederich, Mathias Frisch, David Glick, Jason Grossman, Lucien Hardy, Stephan Hartmann, Richard Healey, Jenann Ismael, Peter Lewis, Gerard Milburn, Jørn Mjelva, Markus Müller, Wayne Myrvold, Heinrich P\"as, Emily Patterson, Carlo Rovelli, Terry Rudolph, Laura Ruetsche, Michael Silberstein, Rob Spekkens, Joan Vaccaro, and Howard Wiseman; apologies to people we have missed. HP is also grateful to audiences at the Perimeter Institute, JHU, Western, IQOQI (Vienna), MCMP (Munich), Hanover, Reykjavik, Dartmouth, and Griffith University, Brisbane.


 

\end{document}